\documentclass[prd,preprint,tightenlines,floatfix,showpacs,preprintnumbers,nofootinbib,eqsecnum]{revtex4}

 \usepackage[dvips,final]{graphicx}
  \usepackage{amssymb}
   \usepackage{amsmath}
    \usepackage{amsfonts}
     \usepackage{epsfig}
      \usepackage{bm}% bold math
\newcommand{\GeV}{\textrm{GeV}}

\begin{document}

\thispagestyle{empty} \preprint{\hbox{}} \vspace*{-10mm}

%\title{Polarisation effects in the central exclusive production
%of $\chi_c$ charmonia \\
%and the $J/\Psi$ decay distribution in the raditive decay channel
%\footnote{Dedicated to the memory of Alexei Kaidalov who passed away
%on July 25th, 2010}}

\title{Polarisation effects in the central exclusive $\chi_c$ production\\
and the $J/\psi$ angular distributions\footnote{Dedicated to the
memory of Alexei Kaidalov who passed away on July 25th, 2010}}

\author{Roman Pasechnik}
 \email{roman.pasechnik@fysast.uu.se}
 \affiliation{Department of Physics and Astronomy,
 Uppsala University, Box 516, SE-751 20 Uppsala, Sweden}

\author{Antoni Szczurek}
 \email{antoni.szczurek@ifj.edu.pl}
 \affiliation{Institute of Nuclear Physics PAN, PL-31-342 Cracow,
 Poland and\\University of Rzesz\'ow, PL-35-959 Rzesz\'ow, Poland}

\author{Oleg Teryaev}
 \email{teryaev@theor.jinr.ru}
 \affiliation{Bogoliubov Laboratory of Theoretical Physics, JINR,
 Dubna 141980, Russia}

\date{\today}

\begin{abstract}
We discuss exclusive elastic double diffractive axial-vector
$\chi_c(1^+)$ and tensor $\chi_c(2^+)$ mesons production for
different meson polarisations in proton-(anti)proton collisions at
the Tevatron energy. The amplitude for the process is derived within the
$k_t$-factorisation approach using unintegrated gluon distributions
(UGDFs). Differential cross sections for different $\chi_c$
polarisations are calculated. Angular distributions of $J/\psi$
meson from the radiative $\chi_c(1^+,2^+)$ decays are derived.
Prospects for experimental selection of different spin states of
$\chi_c$ mesons are discussed.
\end{abstract}

\pacs{13.87.Ce, 13.60.Le, 13.85.Lg}
%Keywords:

\maketitle

%------------------------------
\section{Introduction}
%------------------------------

Recently, the central exclusive production of $\chi_c$ charmonia
has attracted a lot of attention from both experimental
\cite{Albrow:2008pn,Aaltonen:2009kg,Albrow:2010yb} and theoretical
\cite{HarlandLang:2009qe,HarlandLang:2010ep} sides. At the moment,
such a process provides the unique opportunity to test the QCD
diffractive Kaidalov-Khoze-Martin-Ryskin (KKMR) mechanism \cite{KMR}
based on the $k_t$-factorisation incorporating nonperturbative small-$x$
gluon dynamics described by the unintegrated gluon distribution
functions (UGDFs) against accessible data.

The final hadronic central system is, however, rather complicated
and composed of three spin states $\chi_c(0^+,1^+,2^+)$, and there
is no yet any reliable way to measure them separately. Such a
measurement would significantly reduce an overall theoretical
uncertainty, as different $\chi_c$ spin contributions come from
different phase space regions \cite{chic0,chic1,chic2}. Indeed, in
the asymptotical forward limit $\chi_c(1^+,2^+)$ are strongly
suppressed with respect to $\chi_c(0^+)$ due to the $J_z=0$
selections rule \cite{Khoze:2000jm}. However, corrections to this
asymptotics turn out to be important in the total cross section
leading to a noticeable contribution of the $\chi_c(1^+,2^+)$ states
to the observable signal from radiative $\chi_c$ decays
\cite{HarlandLang:2009qe,chic1,chic2}. Generally, separate
measurements of $\chi_c$ states in diffractive production would
impose more strict bounds on the production mechanism under
consideration and extract some new information on the underlying QCD
dynamics \cite{chic2,HarlandLang:2010ep}.

One of the ways is to measure the characteristic differential
distributions like meson $p_{\perp}$ distributions or angular
correlations of the outgoing protons. However, such distributions
are rather sensitive to UGDFs and cuts on kinematical variables
\cite{chic1,chic2}, and
existing experimental techniques do not allow to reconstruct such
observables with sufficient precision. Another possible way is to
look at observables related with meson polarisations and their
decay distributions.

The goal of the present paper is to analyze polarisation effects in
the central exclusive production of $\chi_c$ charmonia. Such effects
can be potentially identified by measuring the angular distribution
of $J/\psi$ mesons from radiative decays of $\chi_c(J^+)$ giving more
detailed information on the partial meson helicity contributions.
Moreover, certain combinations of polarisation observables can be less
sensitive to unknown nonperturbative effects leading to unique
opportunities for model-independent analysis of diffractive
processes.

The paper is organized as follows. In Section II we give general
expressions for the amplitude of the central exclusive
$\chi_c(1^+,2^+)$ production with all necessary notations. Section
III contains discussion of the rapidity dependence of the hard
subprocess amplitudes for different $\chi_c$ polarisations. In
Sections IV and V we derive the angular dependence of outgoing
$J/\psi$ mesons in the helicity frame in terms of the diffractive
$\chi_c$ production density matrix $\rho_{\lambda\lambda'}$.
Sections VI and VII are devoted to discussion of gluon off-shell
effects and absorptive corrections. Section VIII contains the
presentation of the main results, including differential
distributions of polarised $\chi_c(1^+,2^+)$ and angular
correlations of the outgoing $J/\psi$ mesons. Finally, in Section IX
we give a set of concluding remarks and present a discussion of the
final results and theoretical uncertainties.

%----------------------------------------------------
\section{Amplitudes of exclusive P-wave charmonia production}
%----------------------------------------------------

According to the Kaidalov-Khoze-Martin-Ryskin approach (KKMR)
\cite{KMR}, we write the amplitude of the exclusive double
diffractive color singlet production $pp\to pp\chi_{cJ}$ as
\begin{eqnarray}
{\cal
M}^{g^*g^*}_{J,\lambda}=\frac{s}{2}\cdot\pi^2\frac{\delta_{c_1c_2}}{N_c^2-1}\,\Im\int
d^2
q_{0,t}V^{c_1c_2}_{J,\lambda}\frac{f^{\text{off}}_{g,1}(x_1,x_1',q_{0,t}^2,
q_{1,t}^2,t_1)f^{\text{off}}_{g,2}(x_2,x_2',q_{0,t}^2,q_{2,t}^2,t_2)}
{q_{0,t}^2\,q_{1,t}^2\, q_{2,t}^2} \; , \label{ampl}
\end{eqnarray}
where $f^{\text{off}}_{g,i}(x_i,x_i',q_{0,t}^2,q_{i,t}^2,t_i)$ are
the off-diagonal unintegrated gluon distributions for ``active''
gluons with momenta $q_i=x_ip_i+q_{i,t}$ and color indices $c_i$,
and the screening soft gluon with small fraction $x'\ll
x_i,\,q_0\simeq q_{0,t}$, $J$ and $\lambda$ are the spin and
helicity of a produced meson with momentum $P=q_1+q_2$ and mass $M$
in the center-of-mass frame of colliding protons and $z$ axis
directed along the meson momentum ${\bf P}$, respectively.
$V^{c_1c_2}_{J,\lambda}$ here is the hard $g^*g^*\to\chi_{cJ}$
production amplitude for charmonium with quantum numbers $J,\,\lambda$.

The structure of $V_J$ in Eq.~(\ref{ampl})
(we omit here the color and polarisation indices for simplicity) is
determined by gauge invariant amplitude $V_J^{\mu\nu}$ for the
off-shell gluon fusion process, and given in terms of its projection
onto the gluon polarisation vectors as \cite{chic0,chic1,chic2}
\begin{eqnarray}
V_J=n^+_{\mu}n^-_{\nu}V_J^{\mu\nu}=\frac{4}{s}\frac{q^{\nu}_{1,t}}{x_1}
\frac{q^{\mu}_{2,t}}{x_2}V_{J,\mu\nu},\quad
q_1^{\nu}V_J^{\mu\nu}=q_2^{\mu}V_{J,\mu\nu}=0\,.
\end{eqnarray}

In order to define the polarisation states of $\chi_c(1^+,2^+)$
mesons, one has to fix a specific frame, in which those are uniquely
identified and can be measured experimentally. The commonly used
polarisation frames are the helicity, target, Gottfried-Jackson and
Collins-Soper frames (see, e.g. \cite{Lam-Tung}). In this work we
primarily focus on polarisation effects in the helicity frame.

According to Refs.~\cite{chic1,chic2}, in the considered frame we
introduce the time-like basis vectors $n_{1,2,3}$ satisfying
$n^{\mu}_{\alpha}n^{\nu}_{\beta}g_{\mu\nu}=g_{\alpha\beta}$ (with
$n^{\mu}_0=P_{\mu}/M$) with collinear ${\bf n}_3$ and ${\bf P}$
vectors (so, we have ${\bf P}=(E,0,0,P_z),\;P_z=|{\bf P}|>0$), and
\begin{eqnarray}
n_1^{\beta}=(0,\,1,\,0,\,0),\;\; n_2^{\beta}=(0,\,0,\,1,\,0),\;\;
n_3^{\beta}=\frac{1}{M}\,(|{\bf P}|,0,0,E),\;\; |{\bf
P}|=\sqrt{E^2-M^2}. \label{basis}
\end{eqnarray}
Then the hard production amplitudes for the axial-vector $J=1$
($\lambda=0,\,\pm1$) and tensor $J=2$ ($\lambda=0,\,\pm1,\,\pm2$)
charmonia in the helicity frame read (for more details, see
Refs.~\cite{chic1,chic2})
\begin{eqnarray}\nonumber
V^{c_1c_2}_{J=1,\,\lambda}&=&-8g^2\delta^{c_1c_2}\sqrt{\frac{6}{M\pi
N_c}} \frac{{\cal R}'(0)}{|{\bf P}_t|(M^2-q_{1,t}^2-q_{2,t}^2)^2}
\biggl\{\frac{1}{\sqrt{2}}\biggl[i|\lambda|(q_{1,t}^2-q_{2,t}^2)(q_{1,t}q_{2,t})
\mathrm{sign}(\sin\psi)+\\
&&\lambda(q_{1,t}^2+q_{2,t}^2)\,|[{\bf q}_{1,t}\times{\bf
q}_{2,t}]\times{\bf
n}_1|\,\mathrm{sign}(Q^y)\,\mathrm{sign}(\cos\psi)\biggr]+\label{V-fin-chic1}\\
&&(1-|\lambda|)(q_{1,t}^2+q_{2,t}^2)\,|[{\bf q}_{1,t}\times{\bf
q}_{2,t}]\times{\bf
n}_3|\,\mathrm{sign}(Q^y)\,\mathrm{sign}(\sin\psi)\biggr\},\nonumber
\end{eqnarray}
\begin{eqnarray}
V^{c_1c_2}_{J=2,\,\lambda}&=&2ig^2\delta^{c_1c_2}\sqrt{\frac{1}{3M\pi
N_c}}\frac{{\cal R}'(0)}{M|{\bf
P}_t|^2(M^2-q_{1,t}^2-q_{2,t}^2)^2}\times\label{V-fin-chic2}\\
&&\biggl[6M^2\,i|\lambda|(q_{1,t}^2-q_{2,t}^2)\,\mathrm{sign}(Q^y)\Big\{|[{\bf
q}_{1,t}\times{\bf q}_{2,t}]\times{\bf
n}_1|\,(1-|\lambda|)\,\mathrm{sign}(\sin\psi)\,\mathrm{sign}(\cos\psi)+\nonumber\\
&&2\,|[{\bf q}_{1,t}\times{\bf q}_{2,t}]\times{\bf
n}_3|\,(2-|\lambda|)\Big\}-\big[2q_{1,t}^2q_{2,t}^2+(q_{1,t}^2+q_{2,t}^2)(q_{1,t}q_{2,t})\big]\times\nonumber\\
&&\Big\{3M^2(\cos^2\psi+1)\lambda(1-|\lambda|)+6ME\sin(2\psi)\,\lambda(2-|\lambda|)\,
\mathrm{sign}(\sin\psi)\,\mathrm{sign}(\cos\psi)+\nonumber\\
&&\sqrt{6}\,(M^2+2E^2)\sin^2\psi\,(1-|\lambda|)(2-|\lambda|)\Big\}\biggr]
\; ,
\nonumber
\end{eqnarray}
where $\pm Q^y$ are the $y$-components of the gluon transverse
momenta $q_{1/2,t}$ in considered coordinates, $\psi=[0\,...\,\pi]$
is the polar angle between ${\bf P}$ and the c.m.s. beam axis, and
\begin{eqnarray*}
&&|[{\bf q}_{1,t}\times{\bf q}_{2,t}]\times{\bf
n}_1|=\sqrt{q_{1,t}^2q_{2,t}^2-(q_{1,t}q_{2,t})^2}\,|\cos\psi|,\\
&&|[{\bf q}_{1,t}\times{\bf q}_{2,t}]\times{\bf
n}_3|=\frac{E}{M}\sqrt{q_{1,t}^2q_{2,t}^2-(q_{1,t}q_{2,t})^2}\,|\sin\psi|,\\
&&|{\bf P}_t|^2=|{\bf q}_{1,t}|^2+ |{\bf q}_{2,t}|^2+2|{\bf
q}_{1,t}||{\bf q}_{2,t}|\cos\phi,\\
\end{eqnarray*}
where $\phi$ is the angle between fusing gluons. Amplitudes
(\ref{V-fin-chic1}) and (\ref{V-fin-chic2}) explicitly obey gauge
invariance and Bose symmetry properties with respect to the gluon
momenta interchange $q_1\leftrightarrow q_2$.

Another important feature is that the amplitude of axial-vector
charmonia production turns to zero for on-shell gluons, i.e. when
$q_{1,t}^2=0,\,q_{2,t}^2=0$ due to the Landau-Yang theorem (see,
e.g. Ref.~\cite{chic1}). The gluon virtualities (transverse momenta)
provide a leading effect in the diffractive $\chi_c(1^+)$
production, and, therefore, cannot be neglected
\cite{HarlandLang:2009qe,chic1}. This is the striking difference
between the $k_t$-factorisation approach under consideration and
collinear factorisation which forbids production of axial-vector
states \cite{chic1}.

As was demonstrated in Refs.~\cite{chic1,chic2} amplitudes for
diffractive $\chi_c(1^+,2^+)$ turn to zero in the forward limit,
i.e. when ${\bf q}_{1,t}=-{\bf q}_{2,t}={\bf q}_{0,t}$, due to
symmetry relations. This is a direct consequence of $J_z=0$
selection rule \cite{Khoze:2000jm} saying that CEP of higher spins
$J=1,\,2,\,...$ is strongly suppressed in the forward limit. Like
the gluon virtualities in the $\chi_c(1^+)$ case, the off-forward
corrections provide a leading effect in the case of
$\chi_c(1^+,2^+)$ production, and cannot be neglected in the
integrated cross section. In particular, they lead to a substantial
contribution of $\chi_c(2^+)$ meson CEP \cite{chic2}, close to that
from $\chi_c(0^+,1^+)$ mesons.

Let us now turn to the discussion of the polarisation effects in
diffractive $1^+,\,2^+$ charmonia production and first start from
analytic investigation of helicity amplitudes.

%---------------------------------------------------------------------
\section{Rapidity dependence of $g^*g^*\to \chi_c(J)$ amplitudes squared}
%---------------------------------------------------------------------

For the purpose of illustration\footnote{Please, note that the
content of this Section aims to analytical illustration of specific
rapidity dependence of the hard subprocess $g^*g^*\to \chi_c$
amplitudes squared, not the diffractive $pp\to pp\chi_c$ amplitude
squared.}, it is interesting to look at the $y$-dependence of the
$g^*g^*\to\chi_c(J=1,2)$ (hard) subprocess amplitudes
(\ref{V-fin-chic1}) and (\ref{V-fin-chic2}) for different meson
helicities $\lambda=0,\pm 1$ and $\lambda=0,\pm 1,\pm 2$,
respectively. It is convenient to express them in terms of the
transverse 3-momenta of fusing off-shell gluons $|{\bf q}_{1,t}|$
and $|{\bf q}_{2,t}|$, and the angle between them $\phi$ in the
center-of-mass frame of colliding nucleons with the $z$-axis fixed
along meson momentum ${\bf P}$. In this case, summing the matrix
element squared $|V^J_{\lambda}|^2$ over meson polarisations
$\lambda$ up to some constant normalization factor $N^J$ we get:
\begin{eqnarray}\nonumber
S^{J=1}&=&N^{J=1}\frac{|{\bf q}_{1,t}|^2|{\bf q}_{2,t}|^2
\Big[\left(|{\bf q}_{1,t}|^2+|{\bf
q}_{1,t}|^2\right)^2\sin^2\phi+M^2(|{\bf q}_{1,t}|^2+ |{\bf
q}_{2,t}|^2-2|{\bf q}_{1,t}||{\bf q}_{2,t}|\cos\phi)\Big]}
{(|{\bf q}_{1,t}|^2+|{\bf q}_{2,t}|^2+M^2)^4},\\
S^{J=2}&=&N^{J=2}\frac{|{\bf q}_{1,t}|^2|{\bf q}_{2,t}|^2
\Big[3M_{\perp}^2M^2+\left(|{\bf q}_{1,t}|^2\cos\phi+|{\bf
q}_{1,t}|^2\cos\phi+2|{\bf q}_{1,t}||{\bf q}_{2,t}|\right)^2\Big]}
{(|{\bf q}_{1,t}|^2+|{\bf q}_{2,t}|^2+M^2)^4} \label{tot-sum}
\end{eqnarray}
These sums are proportional to the ones derived by Kniehl et al. in
Ref.~\cite{Saleev06} (see right below Eq.~(27)). The distinction is
only due to different normalizations of gluon polarisation vectors
$\epsilon_{\mu}=q^{\mu}_{1/2,t}/|{\bf q}_{1/2,t}|$ used in
Ref.~\cite{Saleev06} and light-cone vectors $n_{\mu}^{\pm}$ used in
our calculation. The sums (\ref{tot-sum}) are $y$-independent. Then,
for partial polarisations of the $\chi_c(1^+)$ meson we have
\begin{eqnarray}\nonumber
&&|V^{J=1}_{\lambda=0}|^2=S^{J=1}\frac{{\cal A}M_{\perp}^2\cosh^2y}{(M_{\perp}^2\cosh^2y-M^2)({\cal B}M^2+{\cal A})},\\
&&|V^{J=1}_{\lambda=\pm1}|^2=\frac{S^{J=1}}{2}\frac{M^2\big[{\cal
B}(M_{\perp}^2\cosh^2y-M^2)-{\cal
A}\big]}{(M_{\perp}^2\cosh^2y-M^2)({\cal B}M^2+{\cal A})},
\label{y-dep-chic1}
\end{eqnarray}
where
\begin{eqnarray*}
{\cal A}=\left(|{\bf q}_{1,t}|^2+|{\bf
q}_{1,t}|^2\right)^2\sin^2\phi,\quad {\cal B}=|{\bf q}_{1,t}|^2+
|{\bf q}_{2,t}|^2-2|{\bf q}_{1,t}||{\bf q}_{2,t}|\cos\phi\,.
\end{eqnarray*}
Analogously, for the tensor $\chi_c(2^+)$ meson we have
\begin{eqnarray}\nonumber
&&|V^{J=2}_{\lambda=0}|^2=S^{J=2}\frac{{\cal R}\big[M_{\perp}^4\cosh
4y + 4(M_{\perp}^2+M^2)M_{\perp}^2\cosh
2y+3M_{\perp}^4+4M_{\perp}^2M^2+2M^4\big]}
{{\cal Q}(M_{\perp}^4\cosh 4y+4{\cal M}\cosh 2y+3M_{\perp}^4-8M^2|{\bf P}_{\perp}|^2)},\\
&&|V^{J=2}_{\lambda=\pm1}|^2=\frac{S^{J=2}}{2}\frac{3M_{\perp}^2M^2\big[|{\bf
P}_{\perp}|^2 M_{\perp}^2\cosh4y+4{\cal P}(\cosh2y+1)-|{\bf
P}_{\perp}|^2M_{\perp}^2\big]} {{\cal Q}(M_{\perp}^4\cosh 4y+
4{\cal M}\cosh 2y+3M_{\perp}^4-8M^2|{\bf P}_{\perp}|^2)}\,,\label{y-dep-chic2}\\
&&|V^{J=2}_{\lambda=\pm2}|^2=\frac{S^{J=2}}{2}\frac{3M^4\big[M_{\perp}^4(\cosh4y+1)-4M^2M_{\perp}^2\cosh2y-2{\cal
P}+2M^4\big]} {{\cal Q}(M_{\perp}^4\cosh 4y+ 4{\cal M}\cosh
2y+3M_{\perp}^4-8M^2|{\bf P}_{\perp}|^2)}\,,\nonumber
\end{eqnarray}
where for compactness we have introduced the following short-hand
notations:
\begin{eqnarray*}
&&{\cal R}=\left(|{\bf q}_{1,t}|^2\cos\phi+|{\bf
q}_{1,t}|^2\cos\phi+2|{\bf q}_{1,t}||{\bf q}_{2,t}|\right)^2,\\
&&{\cal Q}=3M_{\perp}^2M^2+{\cal R},\quad {\cal P}=|{\bf
P}_{\perp}|^4-{\cal R},\quad {\cal M}=|{\bf P}_{\perp}|^4-M^4\,.
\end{eqnarray*}
%---------------------------------------
\begin{figure}[!h]
\begin{minipage}{0.45\textwidth}
 \centerline{\includegraphics[width=1.0\textwidth]{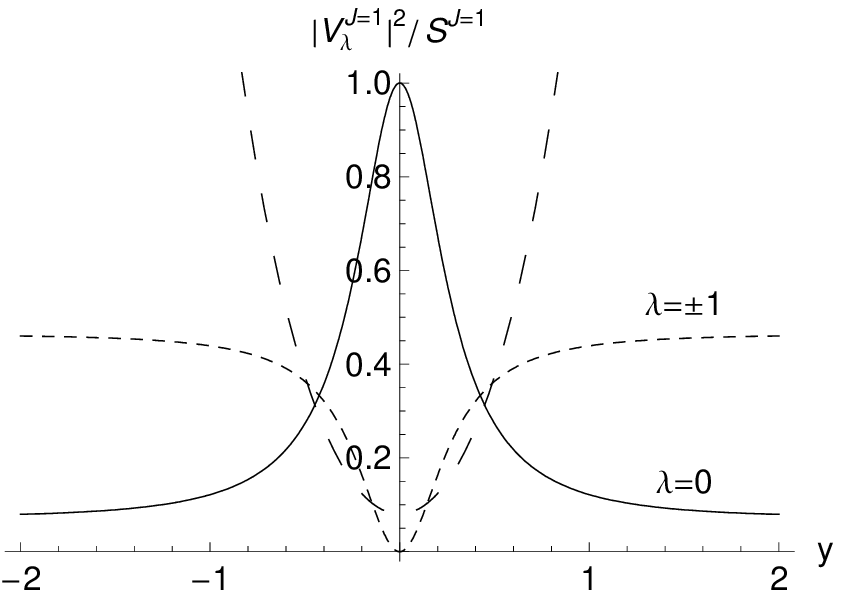}}
\end{minipage}
\hspace{0.5cm}
\begin{minipage}{0.45\textwidth}
 \centerline{\includegraphics[width=1.0\textwidth]{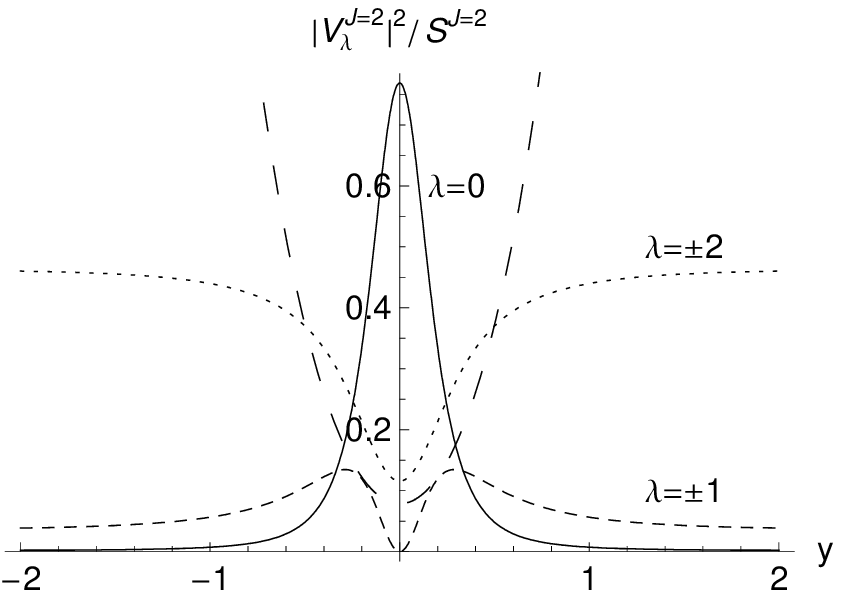}}
\end{minipage}
   \caption{
 \small Illustration of the meson rapidity dependence of the normalized
 subprocess $g^*g^*\to \chi_c(1^+)$ (left panel) and $g^*g^*\to \chi_c(2^+)$
 (right panel) amplitudes squared for different meson helicities $\lambda$
 at arbitrarily fixed kinematical variables
 $|{\bf q}_{1,t}|=|{\bf q}_{1,t}|=0.5$ GeV and $\phi=0.2$.}
 \label{fig:ydep}
\end{figure}
%---------------------------------------

For illustration in Fig.~\ref{fig:ydep} we show the helicity amplitudes
squared as a function of meson rapidity at some fixed kinematical point.
Apparently, they obey a quite non-trivial strong $y$-dependence in
the vicinity of $y=0$ which will result in the ``peaked'' structure
(maxima/minima) of the differential cross sections
$d\sigma^J_{\lambda}/dy$ of double diffractive polarised
$\chi_c(1^+,2^+)$ production (see Results section).

It follows from Eqs.~(\ref{y-dep-chic1}) and (\ref{y-dep-chic2})
that in the forward limit, corresponding to ${\bf q}_{1,t}=-{\bf
q}_{2,t}$, helicity amplitudes $V^{J=1}_{\lambda=0}$ and
$V^{J=2}_{\lambda=0,\pm1}$ turn to zero. So, the total signal is
dominated only by maximal helicity contributions, i.e. by
$\lambda=\pm1$ for $\chi_c(1^+)$ and $\lambda=\pm2$ for
$\chi_c(2^+)$. Therefore, nontrivial ``peaked'' rapidity dependence
of the cross section around $y=0$, as long as absolute values of
non-maximal $\lambda$ contributions, can be served as a measure of
non-forward corrections in the production process.

Let us consider how to measure such
polarisation effects in radiative decay channel $\chi_c(1^+,2^+)\to
J/\psi+\gamma$ and understand how these effects manifest themselves
in differential distributions of outgoing $J/\psi$ meson.

%----------------------------------------------------------
\section{Diffractive $\chi_{cJ}$ production density matrix}
%----------------------------------------------------------

The cross section for the unpolarised $\chi_{cJ}$ production in the
3-body reaction $pp\to pp\chi_{cJ}$, where $J=1,2$, can be written
as
\begin{eqnarray}
\sigma^J_{\chi_c}=
\sum_{\lambda=-J}^{J}\sigma^J_{\lambda\lambda},\qquad
\sigma^J_{\lambda\lambda'}=\sigma^J_{\chi_c}\cdot
\rho^J_{\lambda\lambda}=\frac{1}{2s}\,\int d^{\,3}PS \cdot
d\sigma_{\lambda \lambda'}(\{...\}) \,, \label{prod-cross}
\end{eqnarray}
where $\rho^J_{\lambda\lambda'}$ is the hermitian helicity density
matrix $\rho^{J*}_{\lambda\lambda'}=\rho^J_{\lambda'\lambda}$. In
the differential form it can be written in terms of invariant
diffractive amplitude as
\begin{equation}
d\rho_{\lambda \lambda'}(\{...\})\equiv\frac{d\sigma_{\lambda
\lambda'}}{\sigma^J_{\chi_c}} =\frac{1}{\sigma^J_{\chi_c}}\cdot
\frac{1}{4} \sum_{{\bar \lambda}_1,{\bar \lambda}_2,{\bar
\lambda}'_1,{\bar \lambda}'_2} {\cal M}^*_{{\bar \lambda}_1 {\bar
\lambda}_2 \to {\bar \lambda}'_1 {\bar \lambda}'_2 \lambda}
(\{...\}) {\cal M}_{{\bar \lambda}_1 {\bar \lambda}_2 \to {\bar
\lambda}'_1 {\bar \lambda}'_2 \lambda'} (\{...\}) \; .
\label{production_density_matrix}
\end{equation}
Above $\{...\}$ is an abbreviation for a four-dimensional phase space
point. In our case, because of nucleon helicity conservation (${\bar
\lambda}_1 = {\bar \lambda}_1'$, ${\bar \lambda}_2 = {\bar
\lambda}_2'$), the nucleon helicities can be suppressed. In general,
$\rho^J_{\lambda\lambda'}$ is a function of kinematical variables
$(x_F,t_1,t_2,\phi_{pp})$ (or any other equivalent set of
variables).

Invariance of the amplitude ${\cal M}$ under reflection in
the production plane by the action of the
operator $\exp(-i\pi J_y){\cal P}$, where ${\cal P}$ is the parity
operator, leads to the relation
\begin{eqnarray}
\rho^J_{-\lambda,-\lambda'}=(-1)^{\lambda-\lambda'}\rho^J_{\lambda'\lambda}.
\label{rho-symm}
\end{eqnarray}
Thus, for example, in the case of $J=1$ there are only four independent
components in $\rho^{J}_{\lambda\lambda'}$:
$\rho^1_{00},\;\rho^1_{11},\;\rho^1_{10}$, and $\rho_{-11}$.
The diagonal elements of the density matrix are real.

%------------------------------------------------
\section{Angular distribution of $J/\psi$ meson from $\chi_c(J)$
decay}
%$\chi_c(J) \to J/\psi + \gamma$}
%------------------------------------------------

Let us consider the central exclusive production process $pp \to
pp\chi_{cJ}$ followed by the radiative decay $\chi_{cJ}\to
J/\psi+\gamma$. We shall limit ourselves here to the narrow-width
approximation. Below we follow notations in Ref.~\cite{Kniehl03}.
Let $\theta$ and $\phi$ be the polar and azimuthal angles of the
$J/\psi$ meson in the $\chi_c(J^+)$ rest frame
(this is so-called helicity frame). Then, the
differential cross section of the $J/\psi$ production in the
sequential process $pp \to pp(\chi_{cJ}\to J/\psi+\gamma)$ can be
written as
\begin{eqnarray}
\frac{d\sigma^{J}_{J/\psi}}{d\Omega}=B_J(\chi_{cJ}\to
J/\psi\gamma)\cdot W(\theta,\phi),\qquad
W(\theta,\phi)=\sum^{J}_{\lambda,\lambda'=-J}
\rho^J_{\lambda\lambda'}A^J_{\lambda\lambda'}(\theta,\phi),
\label{jpsi-cross}
\end{eqnarray}
where $d\Omega=d\cos{\theta}d\phi$, $B_J$ denotes the branching
fraction of $\chi_{cJ}\to J/\psi\gamma$, $W(\theta,\phi)$ is the
angular distribution of the $J/\psi$ meson,
$\rho^J_{\lambda\lambda'}$ is the integrated density matrix
corresponding to $\chi_{cJ}$ production process (\ref{prod-cross}),
and $A^J_{\lambda\lambda'}$ refers to the $\chi_{cJ}$ decay process
and allows to describe the decay angular distribution
\begin{eqnarray}
A^J_{\lambda\lambda'}(\theta,\phi)=\frac{\sum_{\lambda_1,\lambda_2}
\langle\lambda_1,\lambda_2,\theta,\phi|T|J,\lambda'\rangle^*
\langle\lambda_1,\lambda_2,\theta,\phi|T|J,\lambda\rangle}
{\int \sum_{\lambda,\lambda_1,\lambda_2}
|T^J_{\lambda,\lambda_1,\lambda_2}(\theta,\phi)|^2 \, d \Omega} \; ,
\end{eqnarray}
where $\lambda_1=0,\pm1$ and $\lambda_2=\pm1$ are the helicities of
the $J/\psi$ and $\gamma$ in the $\chi_{cJ}$ rest frame,
respectively. The transition matrix element of the radiative decay
process $\chi_{cJ}(\lambda)\to
J/\psi(\lambda_1,\theta,\phi)+\gamma(\lambda_2,\pi-\theta,\phi+\pi)$
in the considered frame of reference is \cite{Pilk}
\begin{eqnarray}
&&{}\langle\lambda_1,\lambda_2,\theta,\phi|T|J,\lambda\rangle=
\sqrt{\frac{2J+1}{4\pi}}T^J_{\lambda_1,\lambda_2}
D^{J*}_{\lambda,\lambda_1-\lambda_2}(-\phi,\theta,\phi),\\
&&{}T^J_{\lambda_1,\lambda_2}=\sqrt{\frac{4\pi}{2J+1}}
\langle\lambda_1,\lambda_2,0,0|T|J,\lambda\rangle|_{\lambda=\lambda_1-\lambda_2}
\; ,
\nonumber
\end{eqnarray}
Let $\eta$, $\eta_1$, and $\eta_2$ ($J$, $J_1$, and $J_2$) are the
parities (total angular momenta) of the $\chi_{cJ}$, $J/\psi$, and
$\gamma$, respectively. Angular-momentum conservation imposes the
selection rule $|\lambda_1-\lambda_2|\leq J$. Due to the parity
conservation the decay amplitude $T^J_{\lambda_1,\lambda_2}$
satisfies the following symmetry property \cite{Pilk}
\begin{eqnarray}
T^J_{-\lambda_1,-\lambda_2}=\eta\eta_1\eta_2(-1)^{J_1+J_2-J}T^J_{\lambda_1\lambda_2}
=(-1)^JT^J_{\lambda_1\lambda_2} \; .
\end{eqnarray}
Thus, an independent set of $T^J_{\lambda_1,\lambda_2}$ components
for $J=1,\,2$ reads
\begin{eqnarray}
&&{}J=1:\qquad t^1_0\equiv T^1_{1,1}=-T^1_{-1,-1},\quad
     t^1_1\equiv T^1_{0,-1}=-T^1_{0,1}\,,\\
&&{}J=2:\qquad t^2_0\equiv T^2_{1,1}=T^2_{-1,-1},\quad
     t^2_1\equiv T^2_{0,-1}=T^2_{0,1},\quad t^2_2\equiv T^2_{1,-1}=T^2_{-1,1}\,.
\end{eqnarray}
The matrix
\begin{eqnarray}
D^j_{m'm}(\alpha,\beta,\gamma)=\langle
j,m'|D(\alpha,\beta,\gamma)|j,m\rangle=\exp(-i\gamma
m')d^j_{m'm}(\beta)\exp(-i\alpha m)
\end{eqnarray}
is the representation of the rotation operator
\begin{eqnarray}
D(\alpha,\beta,\gamma)=\exp(-i\gamma J_z)\exp(-i\beta
J_y)\exp(-i\alpha J_x),
\end{eqnarray}
with $\alpha$, $\beta$ and $\gamma$ being the Euler angles, in the
eigenstates $|j,m\rangle$ of $J^2$ and $J_z$. The well-known
$d$-functions $d^j_{m'm}(\beta)=\langle j,m'|\exp(-i\beta
J_y)|j,m\rangle$ may be evaluated from the Wigner's formula
\cite{Sakur}
\begin{eqnarray}\nonumber
&&{}d^j_{m'm}(\beta)=\sum^{\mathrm{min}(j+m,j-m')}_{k=\mathrm{max}(0,m-m')}
(-1)^{k-m+m'}\frac{\sqrt{(j+m)!(j-m)!(j+m')!(j-m')!}}
{k!(k-m+m')!(j+m-k)!(j-m'-k)!}\times\\
&&{}\times\left(\cos\frac{\beta}{2}\right)^{2j+m-m'-2k}\left(\sin\frac{\beta}{2}\right)^{2k-m+m'}  \; .
\end{eqnarray}
Orthogonality condition
\begin{eqnarray}
\int d\Omega D^j_{m'm''}(-\phi,\theta,\phi)D^{j*}_{m
m''}(-\phi,\theta,\phi)=\frac{4\pi}{2j+1}\delta_{m'm}
\end{eqnarray}
provides the normalization of $A^J_{\lambda,\lambda'}$
\begin{eqnarray}
\int d\Omega
A^J_{\lambda\lambda'}(\theta,\phi)=\delta_{\lambda\lambda'},
\end{eqnarray}
so that, upon integration over the solid angle,
Eq.~(\ref{jpsi-cross}) reduces to the narrow-width approximation
formula
\begin{eqnarray}
\sigma^{J}_{J/\psi}=B_J(\chi_{cJ}\to
J/\psi+\gamma)\sigma^J_{\chi_c}.
\end{eqnarray}
Combining all above ingredients together, we get the angular
distribution of the $J/\psi$ meson (\ref{jpsi-cross}) in the general
form
\begin{align}\nonumber
&W^{J=1}(\theta,\phi)=\frac{3\sigma^{J=1}_{\chi_c}}{4\pi}\biggl\{\rho^1_{0,0}
\left[r^1_0\cos^2\theta+\frac{r^1_1}{2}\sin^2\theta\right]+
\rho^1_{1,1}\left[r^1_0\sin^2\theta+
\frac{r^1_1}{2}(1+\cos^2\theta)\right]\\ \nonumber
&\qquad\qquad-\sqrt{2}\sin(2\theta)\left(r^1_0-\frac{r^1_1}{2}\right)
[\mathrm{Re}(\rho^1_{1,0})\cos\phi-\mathrm{Im}(\rho^1_{1,0})\sin\phi]\\
&\qquad\qquad-\sin^2\theta\left(r^1_0-\frac{r_1^1}{2}\right)
[\mathrm{Re}(\rho^1_{1,-1})\cos(2\phi)-\mathrm{Im}(\rho^1_{1,-1})\sin(2\phi)]\biggr\},
\label{jpsi-1}
\end{align}
\begin{align}
&W^{J=2}(\theta,\phi)=\frac{5\sigma^{J=2}_{\chi_c}}{4\pi}\biggl\{\rho^2_{0,0}
\left[\frac14r^2_0(3\cos^2\theta-1)^2+\frac32r^2_1\sin^2\theta\cos^2\theta+
\frac38r^2_2\sin^4\theta\right]\nonumber\\
&+\rho^2_{1,1}\left[3r^2_0\sin^2\theta\cos^2\theta+
\frac{r^2_1}{2}\left(4\cos^4\theta-3\cos^2\theta+1\right)+
\frac{r^2_2}{2}\sin^2\theta(\cos^2\theta+1)\right]\nonumber\\
&+\rho^2_{2,2}\left[\frac34r^2_0\sin^4\theta+
\frac{r^2_1}{2}\sin^2\theta(\cos^2\theta+1)+
\frac{r^2_2}{8}\left(\cos^4\theta+6\cos^2\theta+1\right)\right]\nonumber\\
&+\frac{\sqrt{6}}{4}\sin2\theta\big(2r^2_0(1-3\cos^2\theta)+2r^2_1\cos2\theta+r^2_2\sin^2\theta\big)
\big[\mathrm{Re}(\rho^2_{1,0})\cos\phi-\mathrm{Im}(\rho^2_{1,0})\sin\phi\big]
\nonumber\\
&-\frac12\sin^2\theta\big(6r^2_0\cos^2\theta+r^2_1(\sin^2\theta-3\cos^2\theta)-r^2_2\sin^2\theta\big)
\big[\mathrm{Re}(\rho^2_{1,-1})\cos2\phi-\mathrm{Im}(\rho^2_{1,-1})\sin2\phi\big]\nonumber\\
&+\frac12\sin^3\theta\big(6r^2_0-4r^2_1+r^2_2\big)\Big[
\cos\theta\big[\mathrm{Re}(\rho^2_{2,-1})\cos3\phi-\mathrm{Im}(\rho^2_{2,-1})\sin3\phi\big]\nonumber\\
&\qquad\qquad\phantom{......................}+
\frac14\sin\theta\big[\mathrm{Re}(\rho^2_{2,-2})\cos4\phi-\mathrm{Im}(\rho^2_{2,-2})\sin4\phi\big]\Big]\nonumber\\
&+\frac{\sqrt{6}}{4}\sin^2\theta\big(2r^2_0(3\cos^2\theta-1)-4r^2_1\cos^2\theta+r^2_2(\cos^2\theta+1)\big)
\big[\mathrm{Re}(\rho^2_{2,0})\cos2\phi-\mathrm{Im}(\rho^2_{2,0})\sin2\phi\big]\Big]\nonumber\\
&-\frac14\sin2\theta\big(6r^2_0\sin^2\theta+4r^2_1\cos^2\theta-r^2_2(\cos^2\theta+3)\big)
\big[\mathrm{Re}(\rho^2_{2,1})\cos\phi-\mathrm{Im}(\rho^2_{2,1})\sin\phi\big]\Big]\biggr\}
\; ,
\label{jpsi-2}
\end{align}
where
\begin{eqnarray}
r^J_{\lambda}=\frac{|t^J_{\lambda}|^2}{\sum^J_{\lambda'=0}|t^J_{\lambda'}|^2}
\end{eqnarray}
are positive numbers satisfying $\sum^J_{\lambda=0}r^J_{\lambda}=1$
to be determined experimentally. In Ref.~\cite{e835} these values
were extracted from the Fermilab E835 data on fractional amplitudes
of the electric dipole (E1), magnetic quadrupole (M2), and electric
octupole (E3) transitions in the exclusive reactions
$p\bar{p}\to\chi_{cJ}\to J/\psi\gamma\to e^+e^-\gamma$, with
$J=1,2$. They are listed here
\begin{eqnarray}
&&{}r^1_0=0.498\pm0.032,\qquad r^1_1=0.502\pm0.032,\\
&&{}r^2_0=0.075\pm0.029,\qquad r_1^2=0.250\pm0.048,\qquad
r^2_2=0.674\pm0.052.
\end{eqnarray}
The corresponding values for a pure E1 transitions read
$r_0^1=r_1^1=0.5,\;r_0^2=0.1,\;r_1^2=0.3$, and $r_2^2=0.6$. Having
these values we only need to calculate the independent components of
$\rho^J_{\lambda\lambda'}$ to determine the $J/\psi$ angular
distribution $W^J(\theta,\phi)$ completely.

Analogously to Eq.~(\ref{prod-cross}), we can represent the angular
distribution function as
\begin{eqnarray}
W^{J}(\theta,\phi)=\frac{1}{2s}\,\int  d^{\,3}PS \cdot
dW^J(\theta,\phi;\{...\})\; .
\label{prod-cross-W}
\end{eqnarray}
Such a representation will allow us to calculate correlations
between kinematical variables of the central system as a whole
(for example, rapidity $y$, momentum transfers $t_{1,2}$ and angle
between outgoing protons $\phi_{pp}$) and angular variables of
produced $J/\psi$ meson. Distributions $dW(\theta,\phi;\{...\})$
contain all the information about diffractive $\chi_c$ production
mechanism, meson polarisations and decay process.

%---------------------------------
\section{Gluon off-shell effects}
%---------------------------------

Let us comment now on the importance of gluon virtualities
(transverse momenta) in the hard subprocess vertex
$g^*g^*\to\chi_c(J^+)$ implied by the genuine $k_t$-factorisation
scheme. This question was originally investigated in
Ref.~\cite{chic0} when considering the central exclusive production
of $\chi_c(0^+)$ meson. In particular, it was found that the
on-shell approximation ($M_{\chi_c}^2\gg q_{1,2}^2$) relying on the
neglecting terms of the order of $q_{1,2}^2/M_{\chi_c}^2$ is too
crude for the production of low mass systems with mass being of the
order of a few GeV. In the $\chi_c(0^+)$ CEP case, inclusion of the
gluon virtualities $q_{1,2}^2$ in the hard matrix element leads to a
noticeable decrease of the total (bare) CEP cross section by a
factor of 2 -- 4 depending on the model of UGDF used.

This off-shell effect is even more pronounced in the case of the
axial-vector $\chi_c(1^+)$ production, as it is dominated mostly by
relatively large gluon virtualities and vanishes in the strict
on-shell case. Even when included in the numerator of the vertex
(see, e.g., Eq.~(\ref{V-fin-chic1})) the ``small'' terms $q_{1,2}^2$
are sometimes neglected in its denominator. Such an approximation
$M_{\chi_c}^2\gg q_{1,2}^2$ increases the total $\chi_c(1^+)$ CEP
cross section by almost a factor of 7 in absolute normalization
making a large effect in the $\chi_c(1^+)/\chi_c(2^+)$ ratio in the
observable radiative decay channel \cite{chic2} (see also Table
\ref{table:dy0} below). This resulted in a large discrepancy (in
both normalisation of the total cross sections and relative spin
contributions into observable signal) between our previous results
in Refs.~\cite{chic0,chic1,chic2}, where the exact hard matrix
elements with off-shell gluons were used, and HKRS results in
Refs.~\cite{HarlandLang:2009qe,HarlandLang:2010ep}, where gluon
on-shell approximation $M_{\chi_c}^2\gg q_{1,2}^2$ was
adopted\footnote{We are grateful to L. Harland-Lang for the
significant help in understanding the major part of discrepancies
between our calculations.}. Unlike the absolute normalisation of the
cross section, the shapes of the differential distributions are not
strongly affected by such an approximation.

One should note here that the description of the Tevatron data on
the central exclusive $\chi_c$ production with updated results,
including off-shell effects, becomes more difficult as the total
cross section in the radiative decay channel $\chi_c(J^+)\to
J/\psi+\gamma$ gets underestimated by about a factor of two with
respect to the Tevatron data \cite{chic2}, if one uses values for
the gap survival factors calculated in
Refs.~\cite{HarlandLang:2009qe,HarlandLang:2010ep}. However, such
results are still within an overall theoretical uncertainty from the
soft physics involved, UGDF models in the QCD mechanism under
consideration \cite{chic2}, and also from unknown NLO corrections to
the $g^*g^*\to\chi_c(1^+)$ vertex, which are implicitly, but only
partly, taken into account by the $k_t$-factorisation approach used.

%---------------------------------------
\section{Soft rescattering corrections}
%---------------------------------------

In Refs.~\cite{chic0,chic1,chic2}, for simplicity the whole analysis
of observables has been performed adopting an approximation when all
soft rescattering effects are effectively absorbed into the
effective gap survival factor $\langle S^2_{\mathrm{eff}} \rangle$
taken into account multiplicatively. However, it is known from
earlier analysis of the central exclusive Higgs production
\cite{KMR} (and later for unpolarized $\chi_c$ production
\cite{HarlandLang:2010ep}) that the soft rescattering may change
angular distributions crucially, so it is not enough to include the
gap survival factor as a constant normalization of the cross
section.

In order to get some insight on the soft rescattering effects and
include them dynamically, in the first approximation we use the
simple one-channel model for the eikonal part of the survival
probability calculated as
\begin{eqnarray}\label{seik}
S_{\mathrm{eik}}^2({\bf p}_{1,t},{\bf p}_{2,t})=\frac{|{\cal
M}^{bare}({\bf p}_{1,t},{\bf p}_{2,t})+{\cal M}^{res}({\bf
p}_{1,t},{\bf p}_{2,t})|^2}{|{\cal M}^{bare}({\bf p}_{1,t},{\bf
p}_{2,t})|^2}
\end{eqnarray}
where ${\bf p}_{1/2,t}$ are the transverse momenta of the final
protons, ${\cal M}^{bare}({\bf p}_{1,t},{\bf p}_{2,t})$ is the
``bare'' amplitude for the $\chi_c$ CEP, defined in
Eq.~(\ref{ampl}), and ${\cal M}^{res}({\bf p}_{1,t},{\bf p}_{2,t})$
is the rescattering correction which can
be written in the form (see, e.g., Eq.~(13) in
Ref.~\cite{Khoze:2002nf})
\begin{eqnarray}\label{res}
{\cal M}^{res}({\bf p}_{1,t},{\bf p}_{2,t})\simeq
\frac{iM_0(s)}{4\pi s(B+2b)}\exp\left(\frac{b^2|{\bf p}_{1,t}-{\bf
p}_{2,t}|^2}{2(B+2b)}\right)\cdot{\cal M}^{bare}({\bf p}_{1,t},{\bf
p}_{2,t})\,,
\end{eqnarray}
where $\mathrm{Im} M_0(s)=s\sigma^{\mathrm{tot}}_{pp}(s)$ (the real
part is small in the high energy limit), $B$ is the $t$-slope of the
elastic $pp$ differential cross section, and $b\simeq 4\,\GeV^{-2}$
is the $t$-slope of the proton form factor. At the Tevatron energy we
have $\sigma^{\mathrm{tot}}_{pp}\simeq 80$ mb and $B\simeq
17\,\GeV^{-2}$.

In Table \ref{table:dy0} we have collected results for differential cross
section $d\sigma_{\chi_c}/dy(y=0)$ for different spin states, as
well as their contributions to the radiative $J/\psi+\gamma$ decay
channel $d\sigma_{J/\psi\gamma}/dy(y=0)$ at Tevatron. For
illustration, here we use only the perturbatively modeled KMR UGDF
\cite{KMR,MR}, which include the Sudakov form factor and GRV94HO
gluon PDF \cite{GRV}. The results are presented for both versions of
the absorptive corrections -- by using the one-channel approximation
according to the eikonal model\footnote{As for enhanced-diagram correction
factor $S_{\mathrm{enh}}$, we are grateful to L. Harland-Lang for
providing us with a grid for its numerical calculation as a function
of transverse momenta.} (\ref{seik}), (\ref{res}), and in the
factorized form with ${\bf p}_t$-averaged (effective) gap survival
factors $\langle S_{\mathrm{eff}}^2\rangle$ obtained recently in
Ref.~\cite{HarlandLang:2010ep}:
\begin{eqnarray} \langle
 S_{\mathrm{eff}}^2(\chi_c(0^+))\rangle\simeq0.0284,\qquad \langle
 S_{\mathrm{eff}}^2(\chi_c(1^+))\rangle\simeq0.0735,\qquad \langle
 S_{\mathrm{eff}}^2(\chi_c(2^+))\rangle\simeq0.0539\,,
 \label{S2eff}
\end{eqnarray}
where it is assumed that the enhanced suppression factor $\langle
S_{\mathrm{enh}}^2\rangle=0.49$ is the same for all spin states.

%=============================================================================================
\begin{table}[!h]
\caption{\label{table:dy0} \small\sf Differential cross section
$d\sigma_{\chi_c}/dy(y=0)$ (in nb) of the exclusive diffractive
production of $\chi_c(0^+,1^+,2^+)$ mesons and their partial and
total signal in radiative $J/\psi+\gamma$ decay channel
$d\sigma_{J/\psi\gamma}/dy(y=0)$ at Tevatron ($W=1.96$ TeV) for the
KMR UGDF \cite{MR}, different cuts on the transverse momentum of the
gluons in the loop $(q^{cut}_{0,t})^2$, with the on-shell gluons
approximation $M_{\chi_c}^2\gg q_{1,2}^2$ in the hard subprocess
part (denoted as ``on-shell'') and without it (``off-shell''). NLO
correction factors to the hard part $K_{\mathrm{NLO}}$
\cite{qcdcorr} and absorptive corrections in factorized
(\ref{S2eff}) and in one-channel model (\ref{seik}), (\ref{res})
forms are included. Contributions of all polarisations are
incorporated here.}
 {\small
\begin{center}
\begin{tabular}{|c|c||c|c||c|c||c|c||c|c|c||}
\hline
 UGDF & $(q^{cut}_{0,t})^2,$ &\multicolumn{2}{l||}{$\quad\;\;\chi_c(0^+)\quad\quad$}&
  \multicolumn{2}{l||}{$\quad\;\;\chi_c(1^+)\quad\quad$}&
  \multicolumn{2}{l||}{$\quad\;\;\chi_c(2^+)\quad\quad$}& \multicolumn{2}{l|}{$\quad\quad$ratio} & signal \\
\cline{3-10}
 approx. & $\GeV^2$ &$\;\,\frac{d\sigma_{\chi_c}}{dy}\;\,$&$\frac{d\sigma_{J/\psi\gamma}}{dy}$&
 $\;\,\frac{d\sigma_{\chi_c}}{dy}\;\,$&$\frac{d\sigma_{J/\psi\gamma}}{dy}$&
 $\;\,\frac{d\sigma_{\chi_c}}{dy}\;\,$&$\frac{d\sigma_{J/\psi\gamma}}{dy}$&
 $1^+/0^+$
 & $2^+/0^+$ &
 $\sum\frac{d\sigma_{J/\psi\gamma}}{dy}$ \\
\hline\hline
 KMR, on-shell\footnote{In order to avoid a confusion here, please,
note this is not the rigorous on-shell case, but rather an
approximation keeping the leading order term in
$q_{1,2}^2/M_{\chi_c}^2$-expansion.}                           &      &      &      &      &      &      &      &      &      &      \\
 $\langle S_{\mathrm{eff}}^2\rangle$   & 0.72 & 41.5 & 0.47 & 0.51 & 0.17 & 0.30 & 0.06 & 0.36 & 0.13 & 0.70 \\ \hline
 KMR, off-shell                          &      &      &      &      &      &      &      &      &      &      \\
 $\langle S_{\mathrm{eff}}^2\rangle$   & 0.72 & 10.3 & 0.12 & 0.06 & 0.02 & 0.14 & 0.03 & 0.17 & 0.25 & 0.17 \\ \hline
 KMR, off-shell                          &      &      &      &      &      &      &      &      &      &      \\
 $S_{\mathrm{eik}}$ (\ref{res}),
 $S_{\mathrm{enh}}$                    & 0.36 & 29.4 & 0.34 & 0.13 & 0.05 & 0.36 & 0.07 & 0.15 & 0.21 & 0.46 \\ \hline
 KMR, off-shell                          &      &      &      &      &      &      &      &      &      &      \\
 $S_{\mathrm{eik}}$ (\ref{res})
 only                                  & 0.36 & 60.0 & 0.69 & 0.27 & 0.10 & 0.73 & 0.14 & 0.15 & 0.21 & 0.94 \\ \hline
 \hline\hline
 HKRS \cite{HarlandLang:2010ep}        & 0.72 & 35.0 & 0.40 & 0.71 & 0.24 & 0.45 & 0.09 & 0.61 & 0.22 & 0.73 \\
 \hline
\end{tabular}
\end{center}}
\end{table}
%==========================================================================================

With the on-shell approximation $M_{\chi_c}^2\gg q_{1,2}^2$ we basically
reproduce the HKRS observable signal $\sim 0.7$ nb
\cite{HarlandLang:2010ep}. However, we see from Table
\ref{table:dy0} that keeping the exact matrix elements for the hard
subprocess $g^*g^*\to\chi_c(J^+)$ (without the approximation
$M_{\chi_c}^2\gg q_{1,2}^2$ adopted in
Refs.~\cite{HarlandLang:2009qe,HarlandLang:2010ep}) it is rather
difficult to get the Tevatron experimental value
\cite{Aaltonen:2009kg} $d\sigma^{exp}/dy|_{y=0}(pp\to
pp(J/\psi+\gamma))\simeq(0.97\pm0.26)$ nb, even by moving lower cut
on the transverse momentum of the gluons in the loop
$(q^{cut}_{0,t})^2$ down to the minimal perturbative scale
$0.36\,\GeV^2$ in GRV94HO gluon PDF \cite{GRV}. This practically
means that the central exclusive charmonia production is very
sensitive to and seemingly dominated by yet unknown non-perturbative
gluon dynamics at small $x$ and $q_t$.

However, if we exclude the enhanced-diagram correction factor
$S_{\mathrm{enh}}$, which stands for QCD factorisation breaking in
the considered QCD diffractive mechanism, we basically reproduce the
Tevatron data and get the observable signal $\sim 0.94$ nb compared
to the experimental value $0.97$ nb. This may provide a clue that
the QCD factorisation breaking effects may be overestimated and need
further more careful attention.
%====================================================================
\begin{figure}[!h]
\begin{minipage}{0.32\textwidth}
 \centerline{\includegraphics[width=1.3\textwidth]{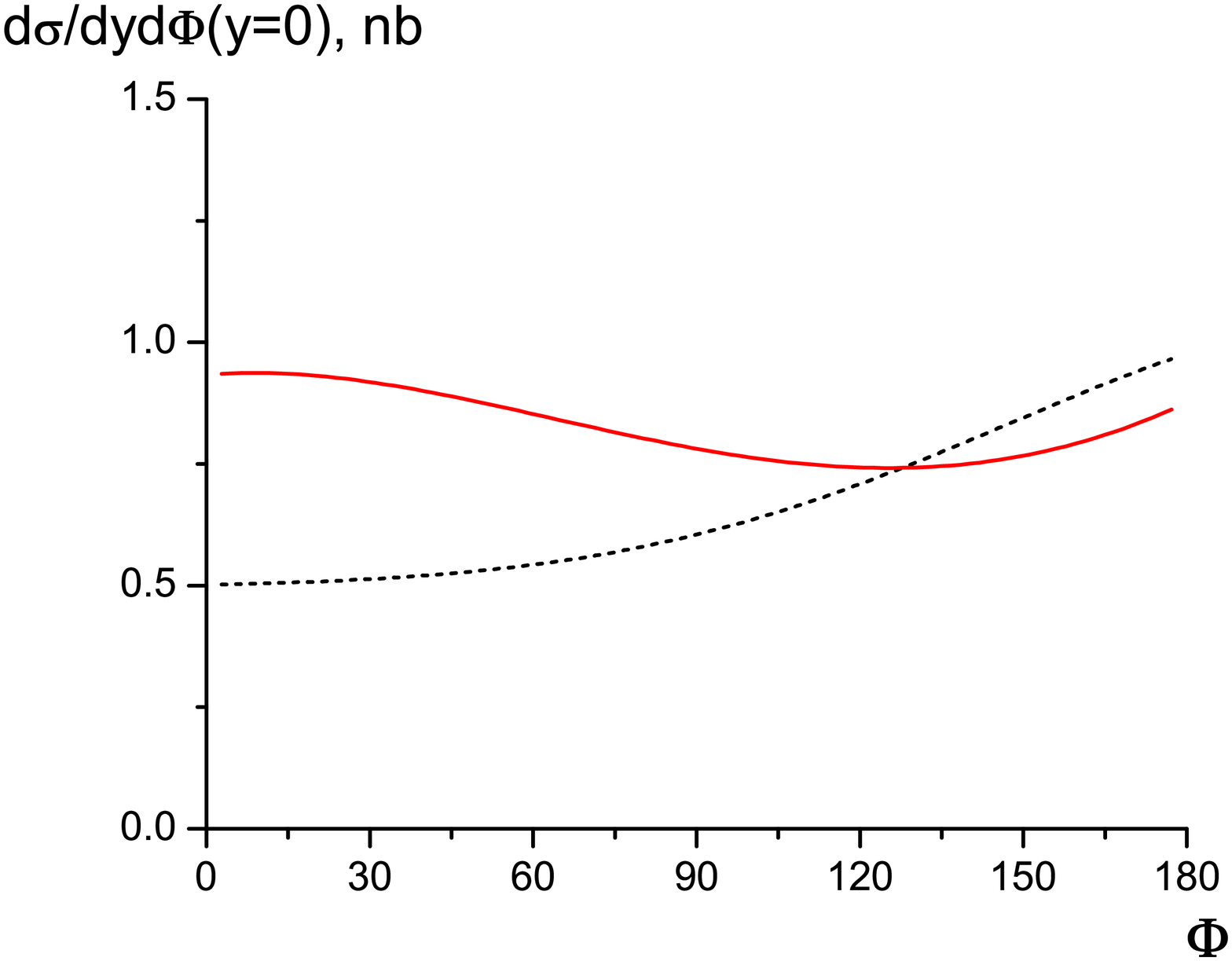}}
\end{minipage}
\begin{minipage}{0.32\textwidth}
 \centerline{\includegraphics[width=1.3\textwidth]{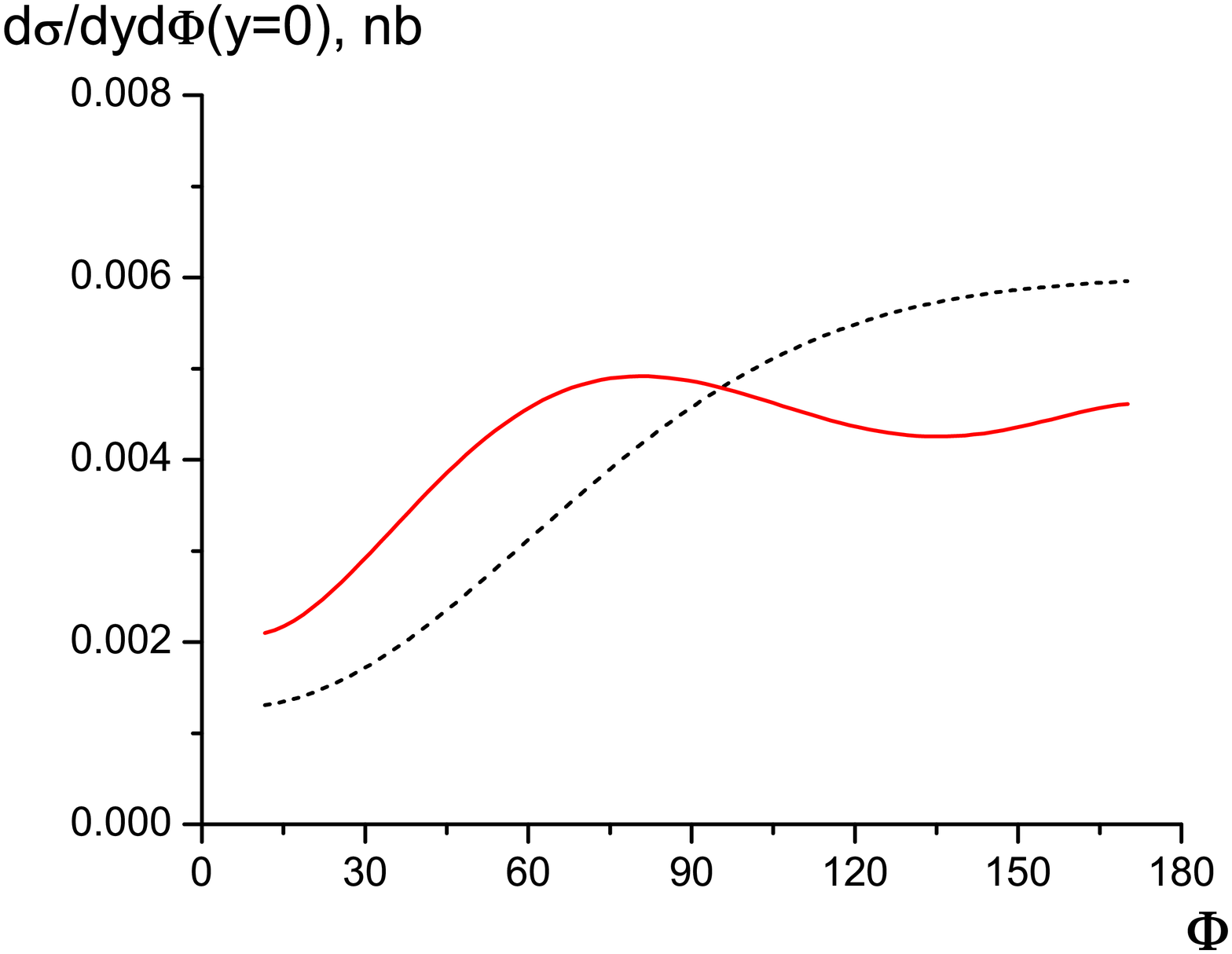}}
\end{minipage}
\begin{minipage}{0.32\textwidth}
 \centerline{\includegraphics[width=1.3\textwidth]{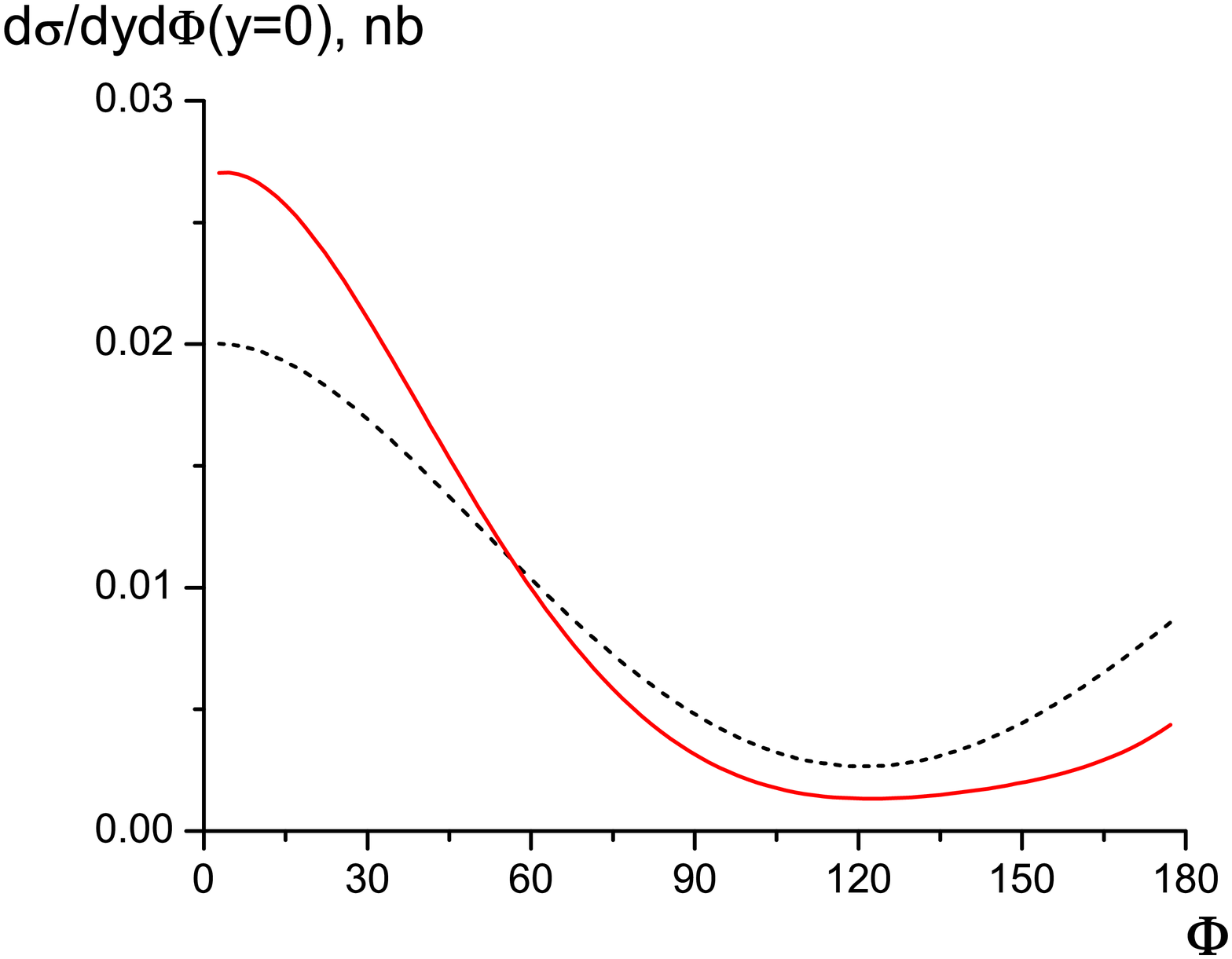}}
\end{minipage}
   \caption{\label{fig:chic-abs}
   \small \em Distributions in relative azimuthal angle
   $\Phi$ between outgoing protons of scalar $\chi_c(0^+)$ (left panel),
   axial-vector $\chi_c(1^+)$ (middle panel)
   and tensor $\chi_c(2^+)$ (right panel) mesons at $y=0$ taking into account
   absorptive corrections in the factorized form (\ref{S2eff}) (dotted lines) and in
   the one-channel model (\ref{seik}), (\ref{res}) (solid lines) forms.
   KMR UGDF \cite{MR} was used.}
\end{figure}
%===================================================================

One-channel eikonal model (\ref{res}) is expected to provide the
main effect on the shape of the distributions in the azimuthal angle $\Phi$
between outgoing protons shown in Fig.~\ref{fig:chic-abs}. Other
distributions in meson rapidity $y$ and momentum transfers along
each proton line $t_{1,2}$ are not changed much -- the effect in
shapes is practically negligible. More general two-channel model
\cite{Khoze:2002nf} may lead to some corrections, but they are not
so big, at least, in the realistic phase space regions, we are
interested in\footnote{We are thankful to L. Harland-Lang
for helpful discussions on this point.}.

%--------------------------------------------------------------
\section{Charmonium polarisation effects in the helicity frame}
%--------------------------------------------------------------

Let us start from presenting differential cross sections. As was
discussed in the previous Section, apart from the overall
normalisation of the total cross section, the absorptive corrections
can noticeably change the shape of distributions in the angle
$\Phi$ between the final protons. As there is no way to measure the
$\Phi$-distributions yet, the measurable effect in $\Phi$-integrated
polarisation observables, which will be discussed below, can be
effectively accounted for by the gap survival factors $\langle
S_{\mathrm{eff}}^2\rangle$ listed in Eq.~(\ref{S2eff}). For
illustration purposes, we show results here only for ``bare''
distributions (without absorptive corrections) calculated with KMR
UGDF \cite{MR}. In principle, absorptive corrections may differ for
different elements of the helicity density matrix, and we leave the
discussion of these effects for a future study.

\begin{figure}[!h]
\begin{minipage}{0.32\textwidth}
 \centerline{\includegraphics[width=1.3\textwidth]{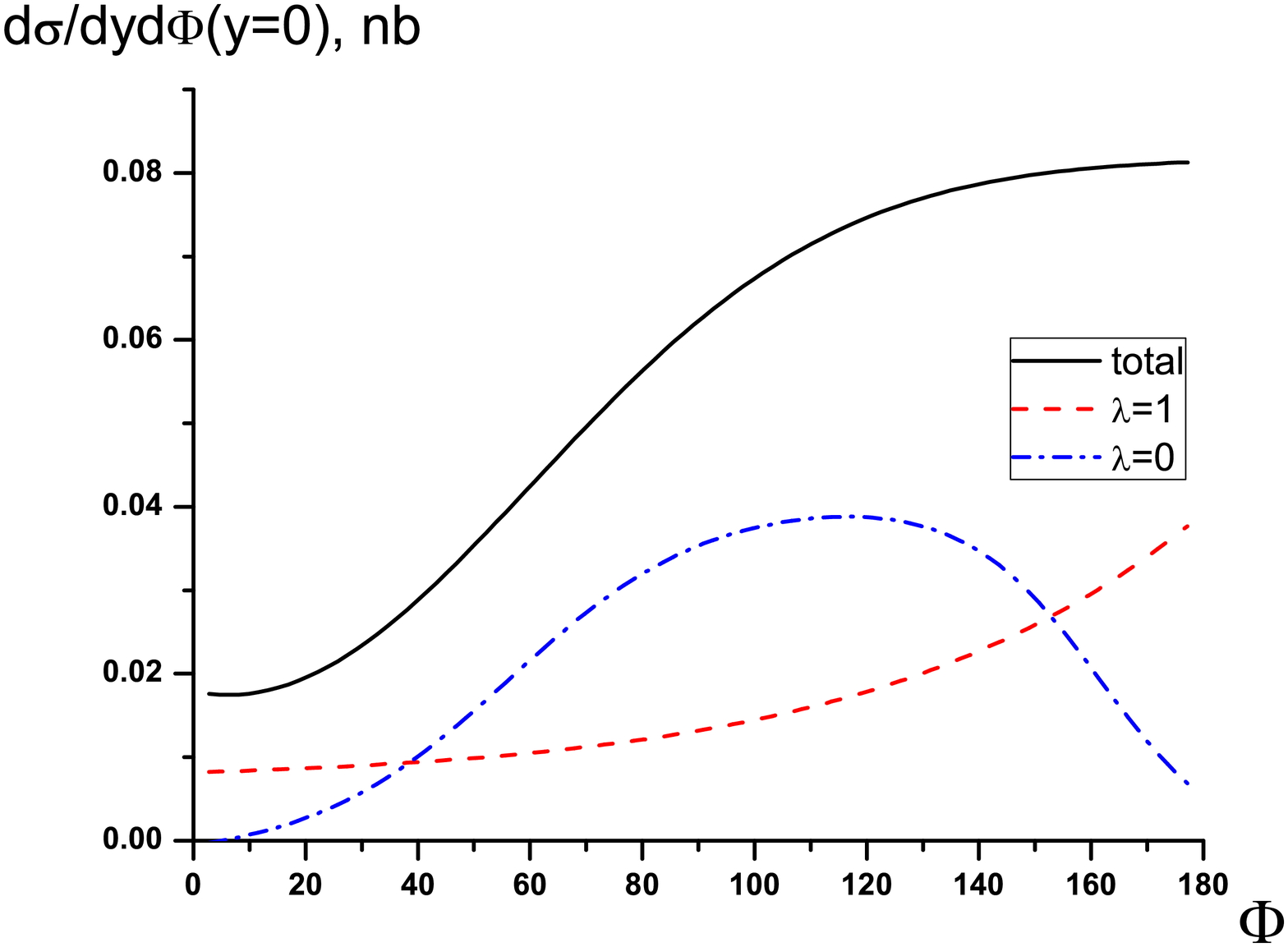}}
\end{minipage}
\begin{minipage}{0.32\textwidth}
 \centerline{\includegraphics[width=1.3\textwidth]{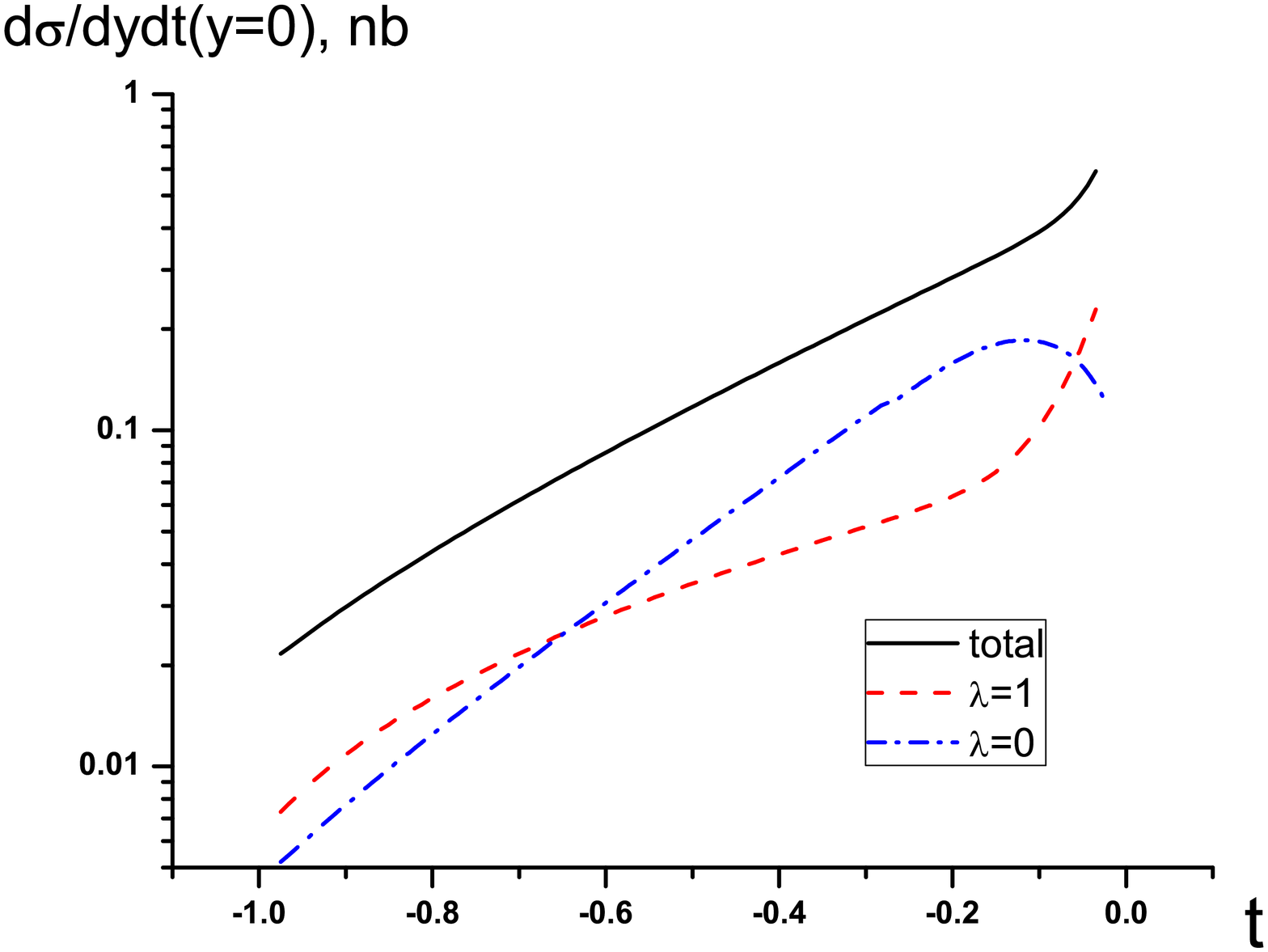}}
\end{minipage}
\begin{minipage}{0.32\textwidth}
 \centerline{\includegraphics[width=1.3\textwidth]{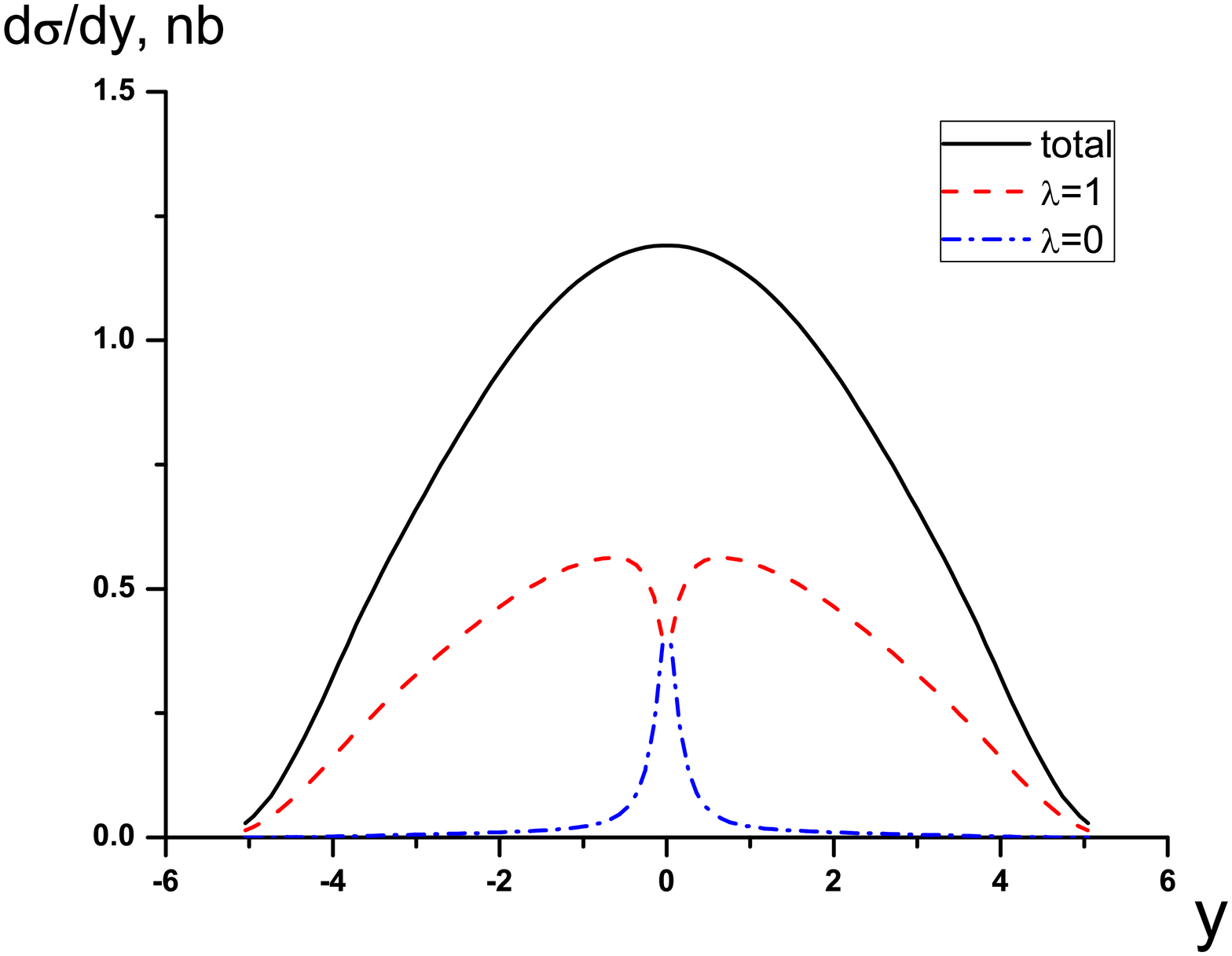}}
\end{minipage}
   \caption{\label{fig:chic1-diag}
   \small \em Distributions of $\chi_c(1^+)$ CEP (bare) cross section in
   relative azimuthal angle $\Phi$ between outgoing protons (left panel),
   momentum transfer $t$ along each proton line (middle panel)
   and meson rapidity $y$ (right panel) for different meson helicities
   $|\lambda|=0,\,1$ (dash-dotted and dashed lines, respectively) and
   for the total (summed over all $\lambda$) CEP cross section (solid line).
   KMR UGDF \cite{MR} was used.}
\end{figure}

\begin{figure}[!h]
\begin{minipage}{0.32\textwidth}
 \centerline{\includegraphics[width=1.3\textwidth]{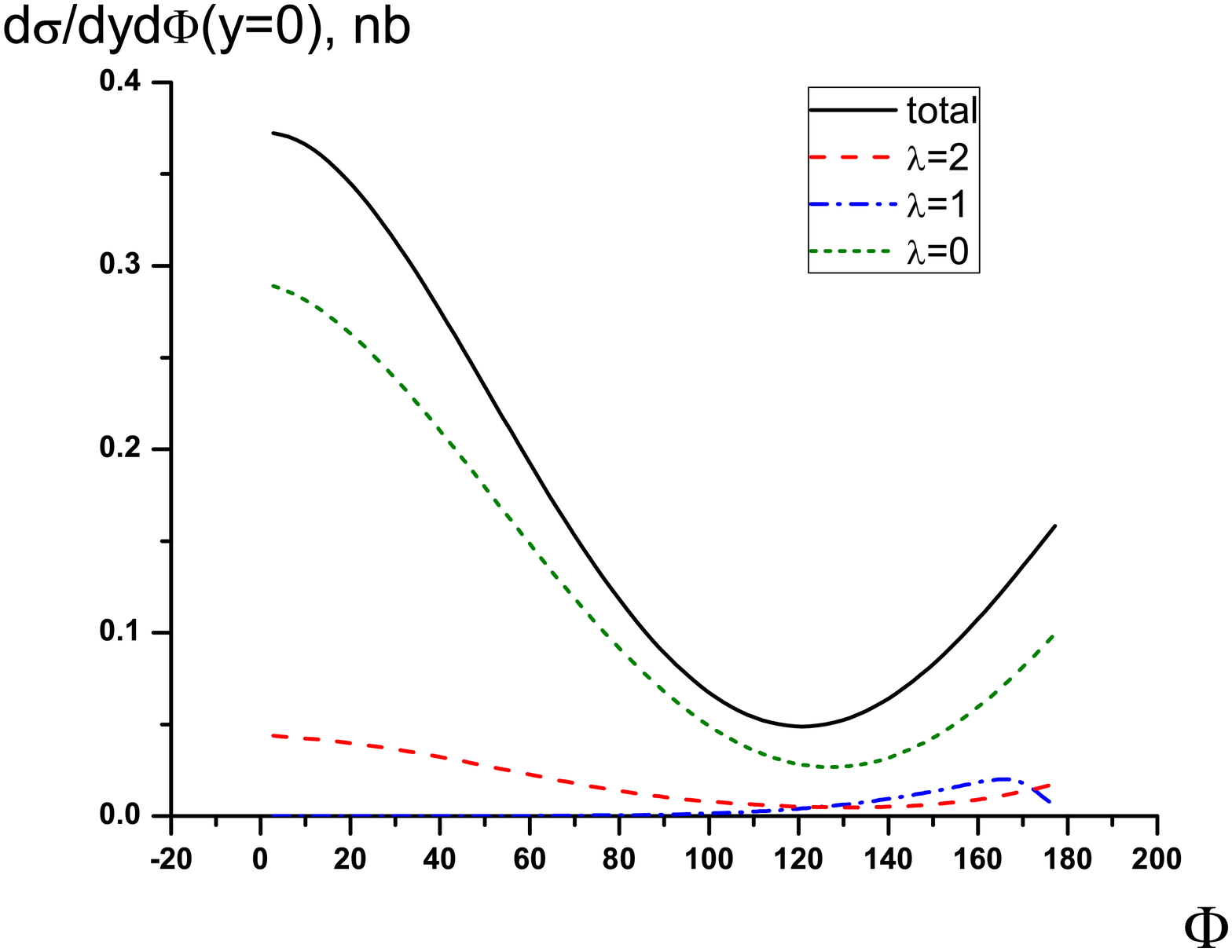}}
\end{minipage}
\begin{minipage}{0.32\textwidth}
 \centerline{\includegraphics[width=1.3\textwidth]{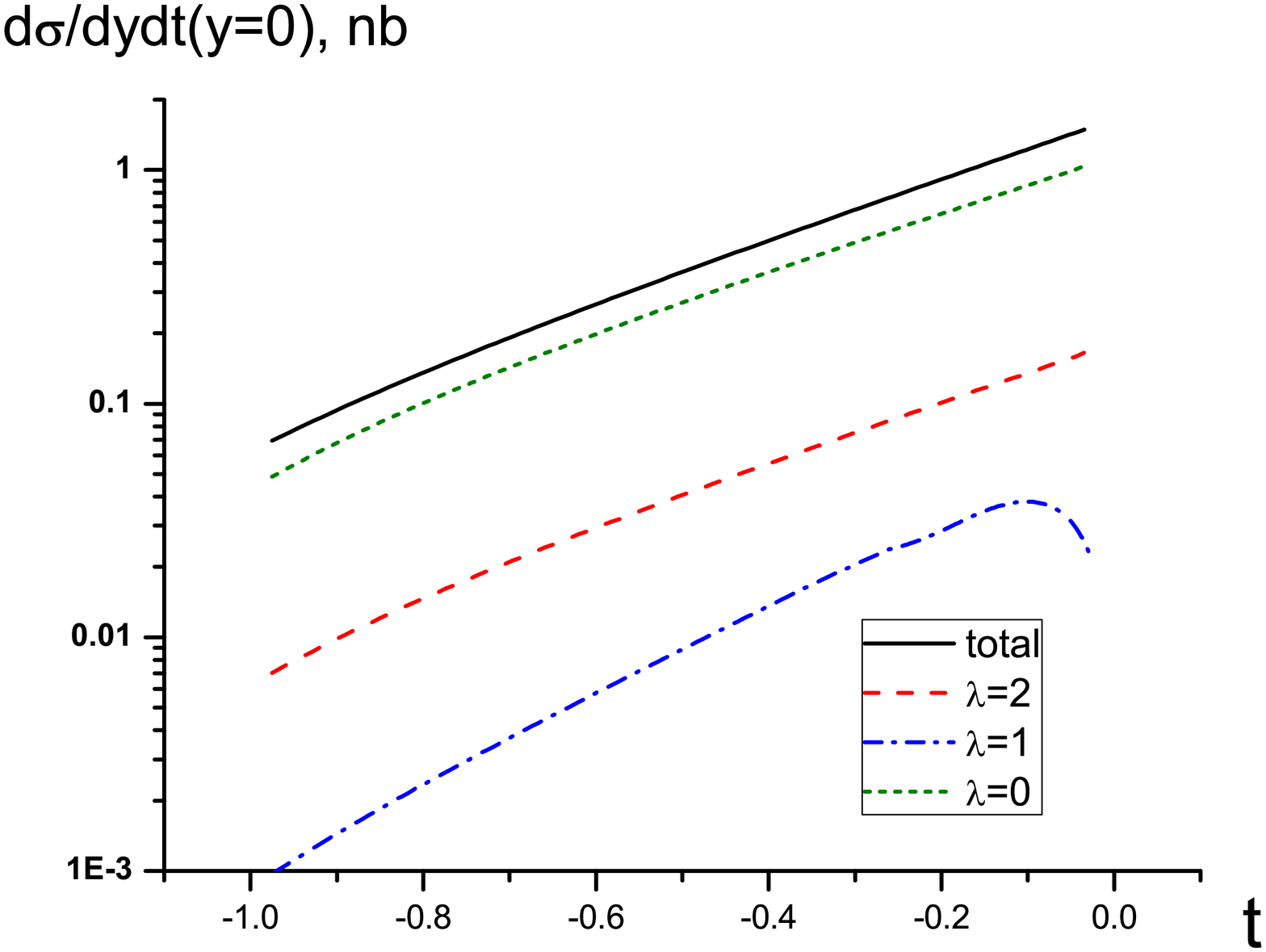}}
\end{minipage}
\begin{minipage}{0.32\textwidth}
 \centerline{\includegraphics[width=1.3\textwidth]{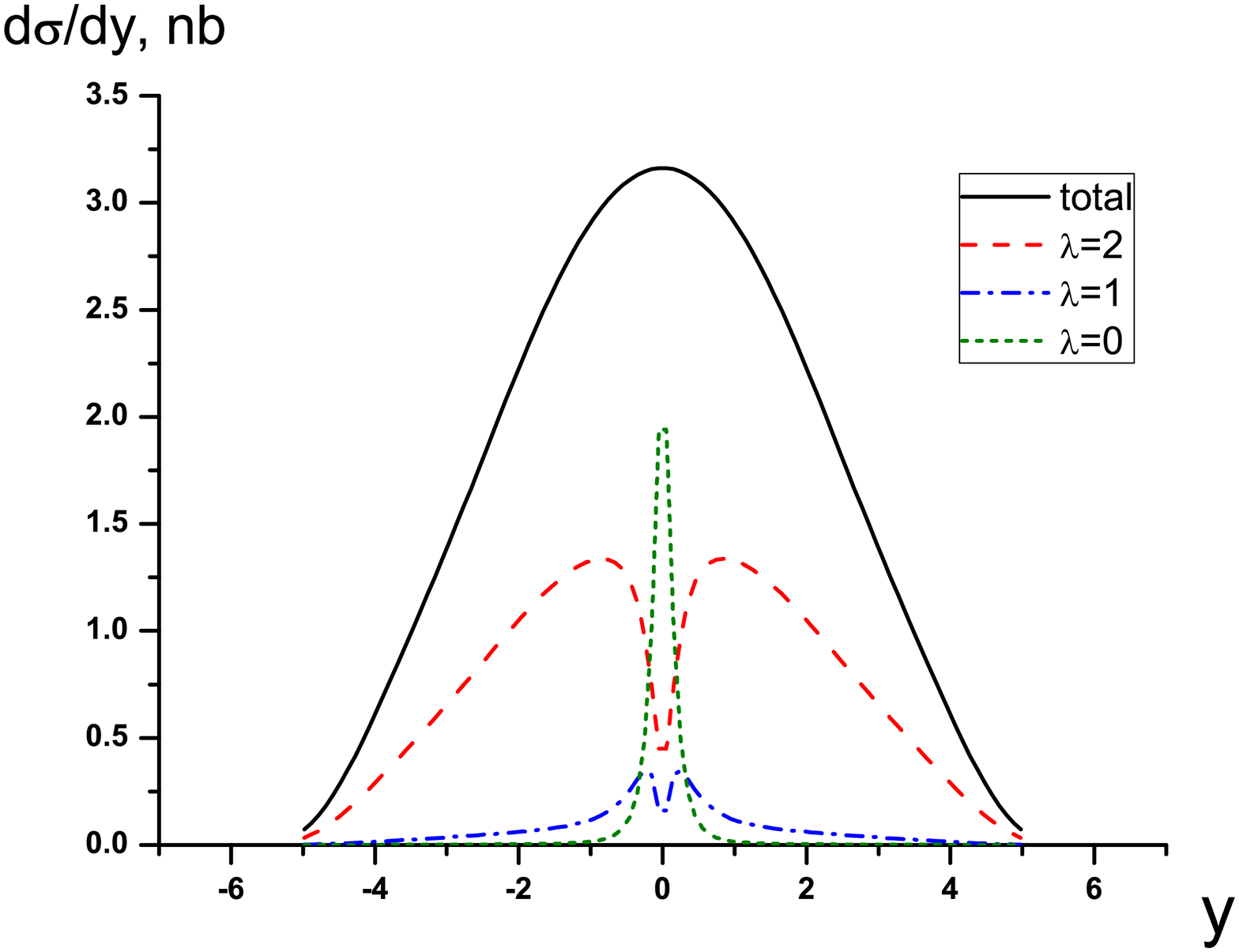}}
\end{minipage}
   \caption{\label{fig:chic2-diag}
   \small \em Distributions of $\chi_c(2^+)$ CEP (bare) cross section in
   relative azimuthal angle $\Phi$ between outgoing protons (left panel),
   momentum transfer $t$ along each proton line (middle panel)
   and meson rapidity $y$ (right panel) for different meson helicities
   $|\lambda|=0,\,1,\,2$ (short-dashed, dash-dotted and dashed lines, respectively)
   and for the total (summed over all $\lambda$) CEP cross section (solid line).
   KMR UGDF \cite{MR} was used.}
\end{figure}

In Fig.~\ref{fig:chic1-diag} we show distributions of the central
exclusive $\chi_c(1^+)$ production cross section in the angle
between outgoing protons $\Phi$, momentum transfer $t$ and meson
rapidity $y$ for different meson polarisations $\lambda =0,\,\pm1$.
In Fig.~\ref{fig:chic2-diag} we present the same distributions for
$\chi_c(2^+)$ meson production. Shapes in $\Phi$ and $t$ are rather
different for both mesons -- at $y=0$ $\chi_c(2^+)$ meson CEP is
always dominated by helicity $\lambda=0$, whereas $\chi_c(1^+)$
production is balanced by both $\lambda=0$ and $\pm1$ contributions,
and which is the dominating one depends on $\Phi$ and $t$ phase
space regions, we look at.
%====================================================================
%===================================================================

The cross sections integrated over all possible meson rapidities $y$
(in our case $|y|\leq 6.0$) are dominated by maximal meson helicity
contributions, i.e. by $|\lambda|=1$ for $\chi_c(1^+)$ and by
$|\lambda|=2$ for $\chi_c(2^+)$ (see Tables~\ref{tab:chic1} and
\ref{tab:chic2} below). In the last case of tensor meson
the $|\lambda|=0$ and $|\lambda|=1$ contributions turn out to have
similar shapes in $\Phi$ and $t$ and the same order of magnitude,
however they are quite different at $y=0$ as it is seen
in Fig.~\ref{fig:chic2-diag}.
%====================================================================
%===================================================================

Differential distributions for fixed meson polarisations
in meson rapidity $y$, shown in
Figs.~\ref{fig:chic1-diag} and \ref{fig:chic2-diag} (right panels),
exhibits maxima/minima in the central rapidity region $y\sim 0$
similar to one presented in Fig.~\ref{fig:ydep}. Interestingly
enough, these maxima/minima in partial helicity contributions cancel
each other in the total (summed over all meson helicity states)
cross section, which has a regular and smooth behavior around $y\to
0$. This is due to a non-trivial $y$-dependence of helicity
amplitudes, which is a pure kinematical effect as discussed in
Section III. Analogously, a non-trivial $y$-dependence of the hard
$g^*g^*\to\chi_c(1^+,2^+)$ subprocess amplitudes squared shown in
Fig.~\ref{fig:ydep} is canceled out in sums over all helicities
$\lambda$ resulting in Eq.~(\ref{tot-sum}). This confirms that the
appearance of non-maximal helicities is a kinematical effect which
is absent in the spin-averaged cross-section.
%-----------------------------------------------------------------
\begin{table}[!h]
\caption{\label{tab:chic1} Non-zeroth integrated elements of the
helicity matrix $\sigma^{J=1}_{\lambda\lambda'},\,\lambda=0,\pm1$
(in nb) for exclusive $\chi_c(1^+)$ production at the Tevatron
energy. KMR \cite{MR} and Kutak-St\'asto UGDFs \cite{KS05} have been
used, absorptive corrections are not included. In the last column
the ratio $\sigma^{J=1}_{00}/\sigma^{J=1}_{\chi_c}=\rho^{J=1}_{00}$
as a measure of subleading helicity-$0$ contribution to the total
cross section is given.}
\begin{center}
\begin{tabular}{|c|c|c|c|c|c|c|}
\hline UGDF
&$\qquad\sigma^{J=1}_{00}\qquad$&$\qquad\sigma^{J=1}_{1,1}\qquad$
&$\qquad\sigma^{J=1}_{1,0}\qquad$&$\qquad\sigma^{J=1}_{1,-1}\qquad$&$\qquad\rho^{J=1}_{00}\qquad$\\
\hline
KMR             & 0.27 & 3.54 & $0.51-0.14i$ & $-1.31+1.50i$  & 0.05  \\
KS              & 0.49 & 3.92 & $0.85-0.08i$ & $-3.14+0.67i$  & 0.06  \\
\hline
\end{tabular}
\end{center}
\end{table}
%------------------------------------------------------------------

Apparently, such a kinematical effect in the CEP cross section
depends on cuts in meson rapidity in a detector. It is a result of
two competing asymptotical effects: heavy (non-relativistic,
$y\to0$) meson is dominantly produced in $\lambda=0$ state, massless
(relativistic, $y\gtrsim 1$) meson is produced mostly in
$\lambda=\pm1$ states. As a result of such a competition, it turns
out that the central exclusive $\chi_c(1^+,2^+)$ production cross
sections are dominated by the second effect such that mostly
relativistic charmonia (with large rapidities $y\gtrsim 1$) in a
maximal helicity state are produced, so the detectors should be
designed to be able to detect them (not only in the central rapidity
region $y\sim0$).
%-----------------------------------------------------------------
\begin{table}[!h]
\caption{\label{tab:chic2} Non-zeroth integrated elements of the
helicity matrix
$\sigma^{J=2}_{\lambda\lambda'},\,\lambda=0,\pm1,\pm2$ (in nb) for
exclusive $\chi_c(2^+)$ production at the Tevatron energy. KMR
\cite{MR} and Kutak-Stasto UGDFs \cite{KS05} have been used,
absorptive corrections are not included. In the last column the
ratio $\sigma^{J=2}_{00}/\sigma^{J=2}_{\chi_c} =\rho^{J=2}_{00}$ as
a measure of subleading helicity-$0$ contribution to the total cross
section is given.}
\begin{center}
\begin{tabular}{|c|c|c|c|c|c|c|c|c|c|c|c|c|c|c|}
\hline UGDF &$\;\sigma^{J=2}_{00}\;$&$\;\sigma^{J=2}_{1,1}\;$
&$\;\sigma^{J=2}_{2,2}\;$&$\;\sigma^{J=2}_{1,0}\;$&$\;\sigma^{J=2}_{2,0}\;$
&$\;\sigma^{J=2}_{1,-1}\;$&$\;\sigma^{J=2}_{1,-2}\;$&$\;\sigma^{J=2}_{1,2}\;$&$\;\sigma^{J=2}_{2,-2}\;$&$\quad\rho^{J=2}_{00}\quad$\\
\hline
KMR    & 0.84 & 0.77 & 7.58 & 0.43 & 0.74 & $-0.54$ & 1.37 & 1.93 & 5.0 & 0.05 \\
KS     & 1.75 & 1.59 & 12.6 & 1.02 & 1.79 & $-1.38$ & 3.17 & 3.95 & 8.3 & 0.06  \\
\hline
\end{tabular}
\end{center}
\end{table}
%-----------------------------------------------------------------

\begin{figure}[!h]
\includegraphics[width=10cm]{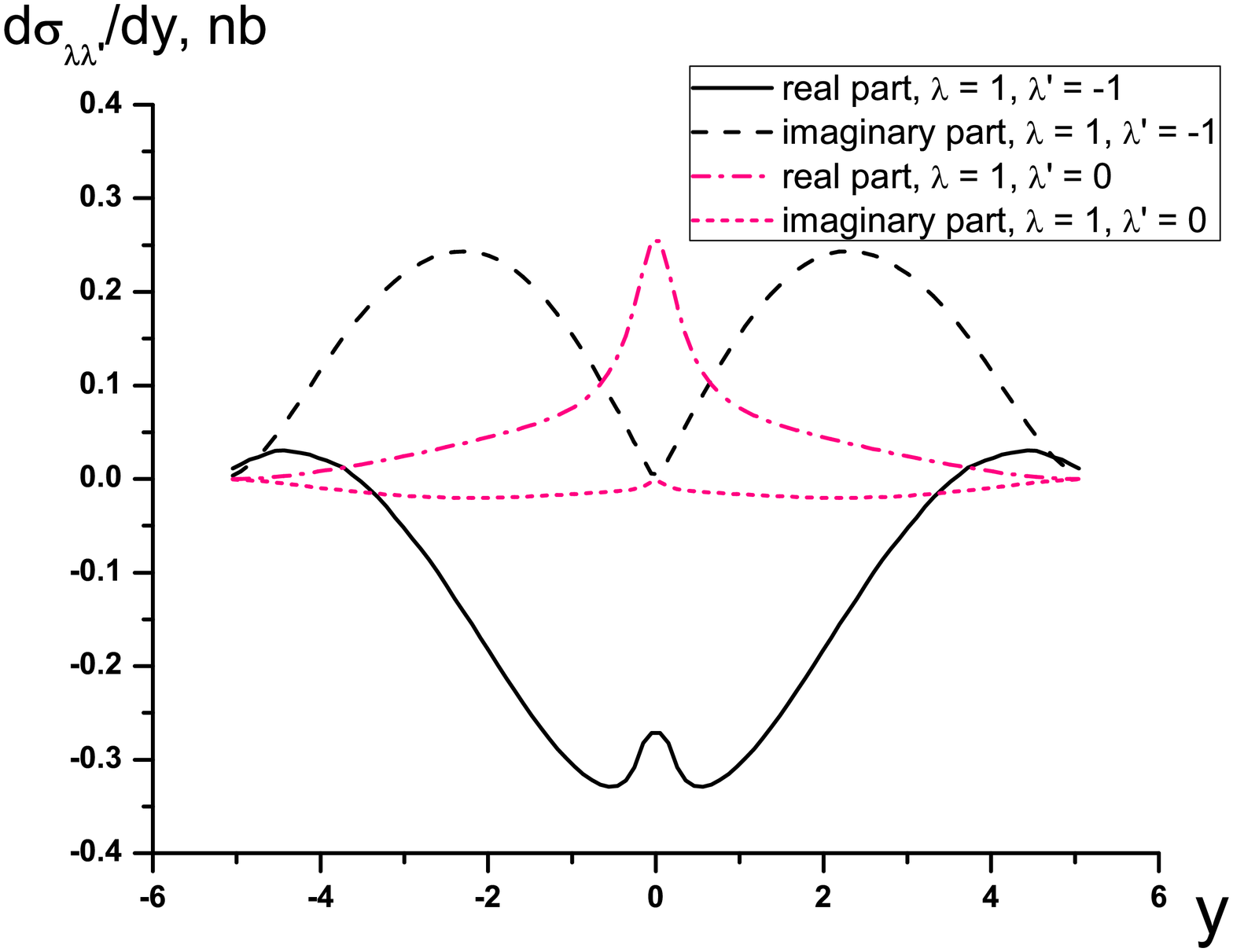}
   \caption{
 \small \em Rapidity distributions of the independent non-diagonal elements
 of the helicity matrix $d\sigma^{J=1}_{\lambda\lambda'}/dy$ (real and imaginary parts)
 for the central exclusive axial-vector $\chi_c(1^+)$ meson production.}
 \label{fig:chic1-ndiag}
\end{figure}

In Tables~\ref{tab:chic1} and \ref{tab:chic2} we present the
integrated elements of the diffractive $\chi_c(1^+,2^+)$ production
helicity matrix
$\sigma^J_{\lambda\lambda'}=\sigma^{J}_{\chi_c}\rho^J_{\lambda\lambda'}$.
Only independent, both diagonal (purely real) and non-diagonal
(complex), non-zeroth elements are listed. Rapidity dependence of
the non-diagonal terms (their real and imaginary parts) in the
axial-vector $J=1$ CEP case is presented for illustration in
Fig.~\ref{fig:chic1-ndiag}.
%-----------------------------------------------------------------
%-----------------------------------------------------------------

As was already pointed out, the total CEP cross section is dominated
by the maximal helicities. The subleading contribution of states
with $\lambda=0$ in the total CEP cross section is about 5-6\% for
both $\chi_c(1^+)$ and $\chi_c(2^+)$ mesons. Measure of such
subleading polarisation contribution is given by the ratio
$\sigma^{J}_{00}/\sigma^{J}_{\chi_c}=\rho^{J}_{00}$ (see, last
columns in Tables~\ref{tab:chic1} and \ref{tab:chic2}). We
calculated these ratios for two typical UGDF models: KMR \cite{MR}
and Kutak-Sta\'sto \cite{KS05}. We see that the ratios
$\rho^{J=1,2}_{00}$ are the same for both mesons, they are almost
independent on UGDF models as the most of theoretical uncertainties
(including soft rescattering effects) are canceled out inside them.
Thus, $\rho^J_{00}$ can be considered as a good model-independent
observable for tests of underlined QCD production mechanism of
$\chi_c$ mesons.

%===================================================================
\begin{figure}[!h]
\begin{minipage}{0.45\textwidth}
 \centerline{\includegraphics[width=1.0\textwidth]{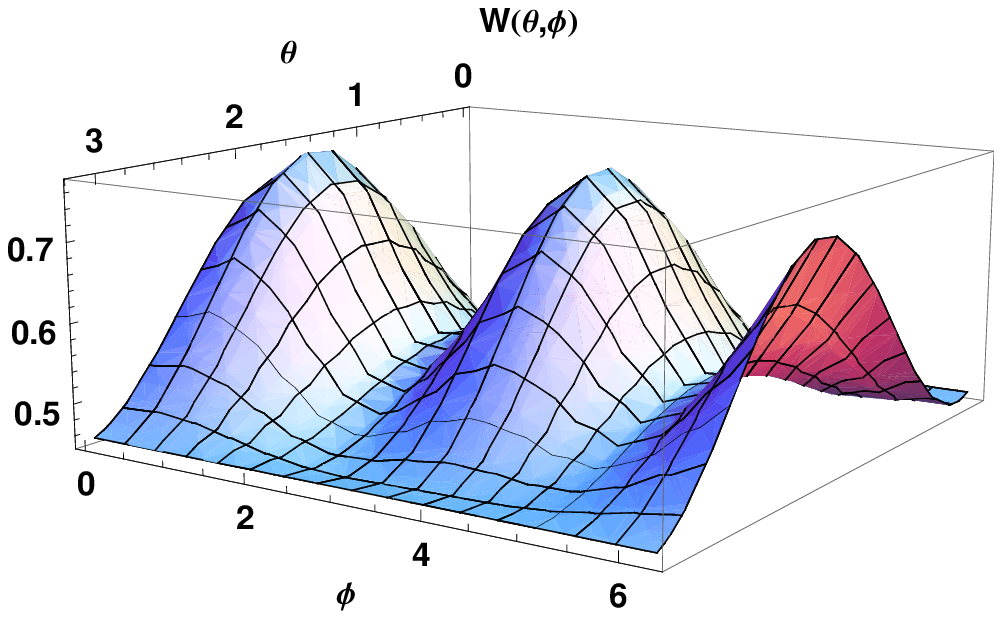}}
\end{minipage}
\begin{minipage}{0.45\textwidth}
 \centerline{\includegraphics[width=1.0\textwidth]{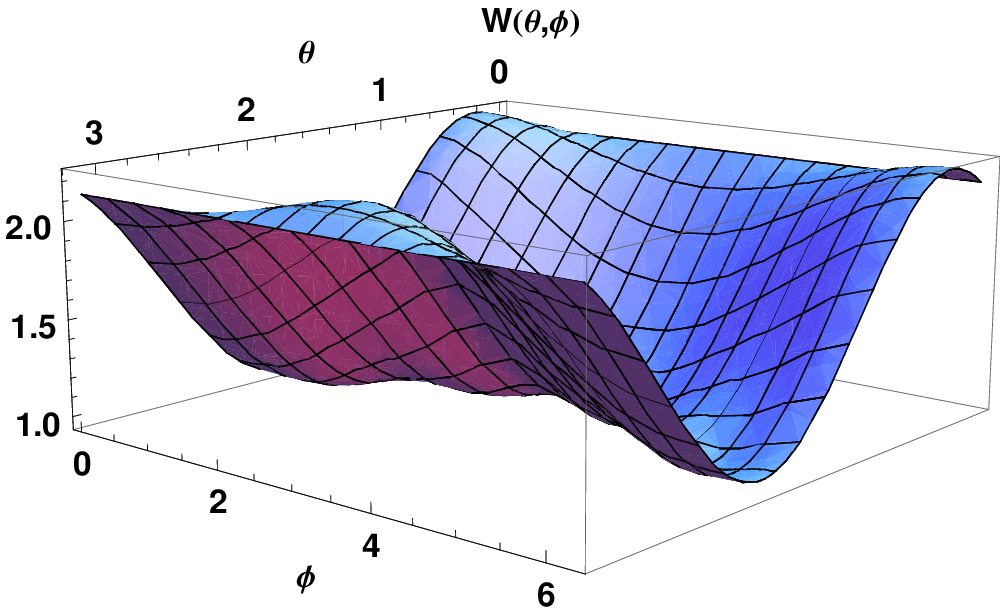}}
\end{minipage}
   \caption{\label{fig:W}
   \small \em Angular distribution $W(\theta,\phi)$ of $J/\psi$ meson
   from radiative decay of centrally produced $\chi_c(1^+)$ (left panel)
   and $\chi_c(2^+)$ (right panel) mesons in polar $\theta$ and
   azimuthal $\phi$ angles in the helicity frame. Full phase space
   for central exclusive $\chi_c$ production is included,
   KMR UGDF \cite{MR} was used.}
\end{figure}
%===================================================================

As it was demonstrated in Ref.~\cite{chic2}, relative contributions
of different spins $J$ are rather sensitive to the model of UGDF.
This is due to the fact that the central exclusive production of
$\chi_c(1^+)$ is more sensitive to small gluon $q_t$ than
$\chi_c(2^+)$ production. Moreover, non-diagonal elements
$\sigma^J_{\lambda\lambda'},\,\lambda\not=\lambda'$ have much
stronger sensitivity to UGDF model, than the diagonal ones.
Therefore, experimental access to different elements of the
production density matrix $\rho^J_{\lambda\lambda'}$ (or
$\sigma^J_{\lambda\lambda'}$) would provide an opportunity for
strong experimental constraints on the unintegrated gluon
distributions at small $x$ and small gluon transverse momenta $q_t$.

Now let us turn to the discussion of observable signal in radiative
decay channel $\chi_c(1^+,2^+)\to J/\psi+\gamma$. As one of the
important characteristics of the production process $pp\to
p(J/\psi\gamma)p$, the angular distribution of the $J/\psi$ meson
$W(\theta,\phi)$ in the helicity frame is shown in Fig.~\ref{fig:W}
as a function of polar $\theta$ and azimuthal $\phi$ angles for both
$\chi_c(1^+,2^+)$ mesons.

%===================================================================
\begin{figure}[!h]
\begin{minipage}{0.45\textwidth}
 \centerline{\includegraphics[width=1.3\textwidth]{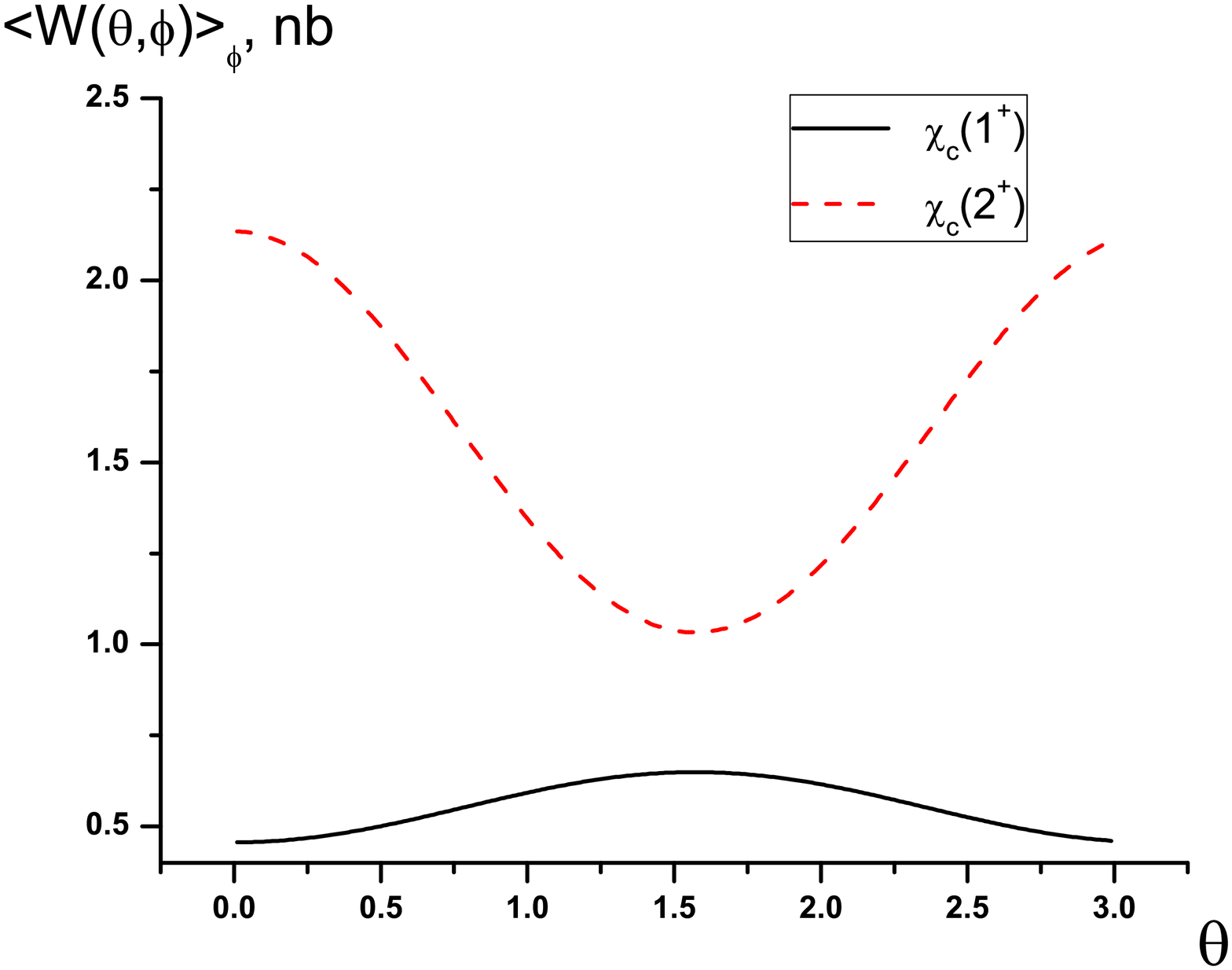}}
\end{minipage}
\begin{minipage}{0.45\textwidth}
 \centerline{\includegraphics[width=1.3\textwidth]{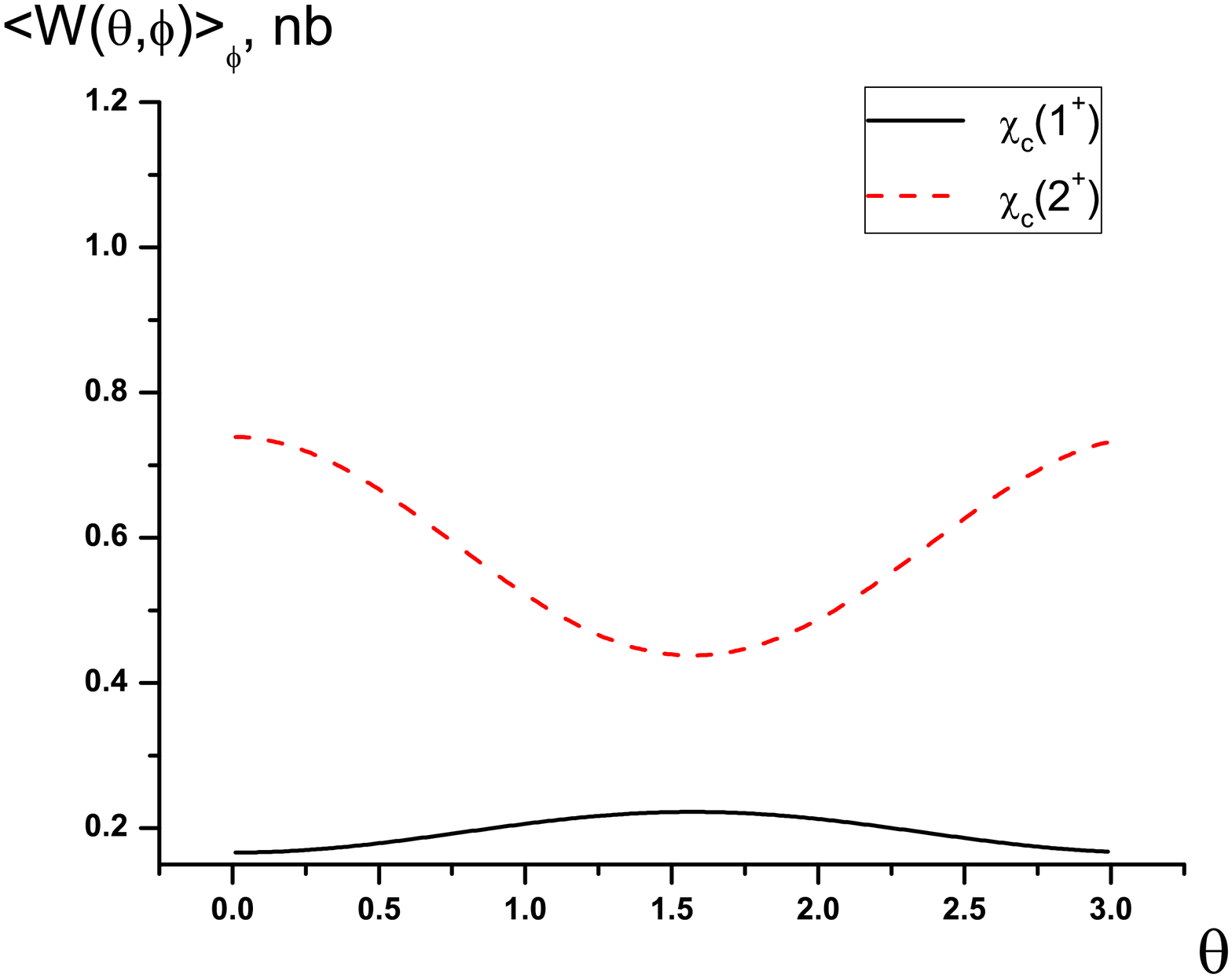}}
\end{minipage}
   \caption{\label{fig:Wav}
   \small \em Angular distribution of $J/\psi$ meson averaged over
 azimuthal angle $\phi$ as function of polar angle $\theta$ (in radians) for
 $\chi_c(1^+)$ meson (solid line) and $\chi_c(2^+)$ meson (dashed line).
 Left panel corresponds to rapidity interval $|y|\leq 6$,
 right panel -- to $|y|\leq 1$.}
\end{figure}
%===================================================================

%-----------------------------------------------------------------
%\begin{figure}[!h]
%\includegraphics[width=8cm]{Wav.eps}
%   \caption{
% \small \em Angular distribution of $J/\psi$ meson averaged over
% azimuthal angle $\phi$ as function of polar angle $\theta$ (in radians) for
% $\chi_c(1^+)$ meson (solid line) and $\chi_c(2^+)$ meson (dashed line). }
% \label{fig:Wav}
%\end{figure}
%-----------------------------------------------------------------

The function $W(\theta,\phi)$ is a periodic one in both the angles
$\theta$ and $\phi$. From Eqs.~(\ref{jpsi-1}) and (\ref{jpsi-2})
it follows that the dependence on polar angle $\theta$ is determined
mostly by the diagonal terms of the production density matrix
$\rho^J_{\lambda\lambda}$, whereas $\phi$-dependence is given by
real and imaginary parts of non-diagonal terms. Moreover, in the
angular distribution $\langle W(\theta,\phi)\rangle_{\phi}$ averaged
over $\phi$ all the terms with non-diagonal elements
$\rho^J_{\lambda\lambda'},\,\lambda\not=\lambda'$ are dropped out
for both $\chi_c(1^+,2^+)$ mesons.

The $\phi$-dependence of $J/\psi$ meson from the $\chi_c(1^+)$ CEP
turned out to be much stronger than that from $\chi_c(2^+)$ CEP as
it is seen from Fig.~\ref{fig:W}. As for the $\theta$-dependence,
periods of oscillations for $\chi_c(1^+)$ and $\chi_c(2^+)$ mesons
are shifted by $\pi/2$ with respect to each other, as demonstrated
in Fig.~\ref{fig:Wav} for distribution $\langle
W(\theta,\phi)\rangle_{\phi}$ averaged over $\phi$. Also, from this
figure we see that the amplitudes of variations in $\theta$ become
smaller and the dominance of the maximal helicity gets weaker and
can be eliminated when one shrinks the rapidity interval in the
phase space integral. This is a direct consequence of the specific
rapidity dependence of the production density matrix elements in the
central rapidity region $y\sim0$.

%===================================================================
\begin{figure}[!h]
\begin{minipage}{0.45\textwidth}
 \centerline{\includegraphics[width=1.0\textwidth]{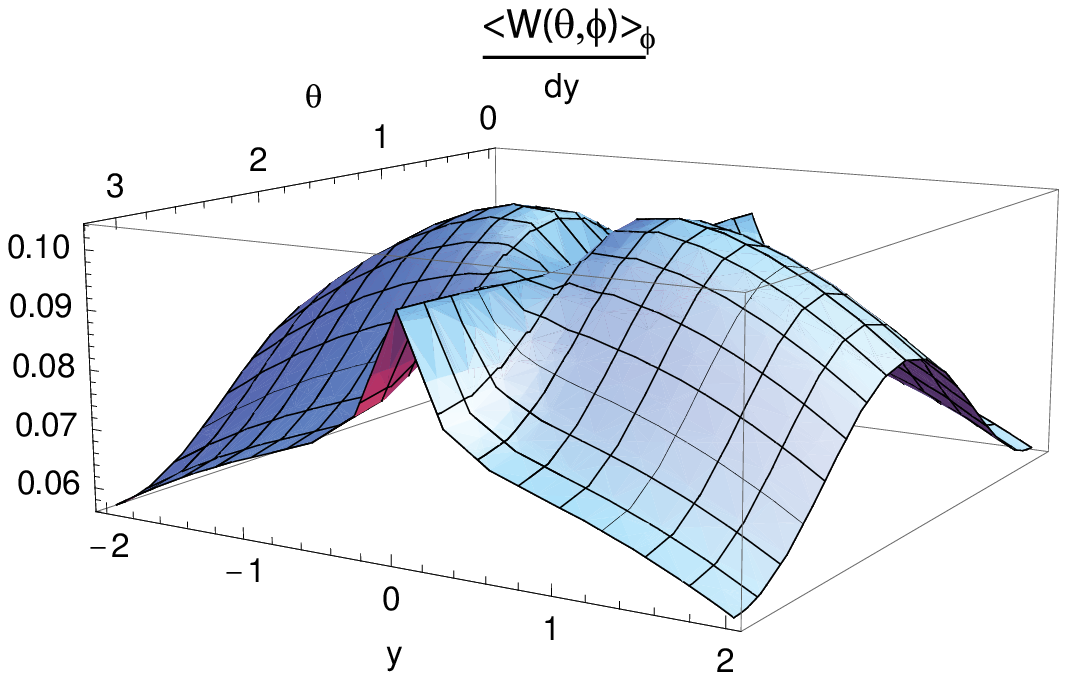}}
\end{minipage}
\begin{minipage}{0.45\textwidth}
 \centerline{\includegraphics[width=1.0\textwidth]{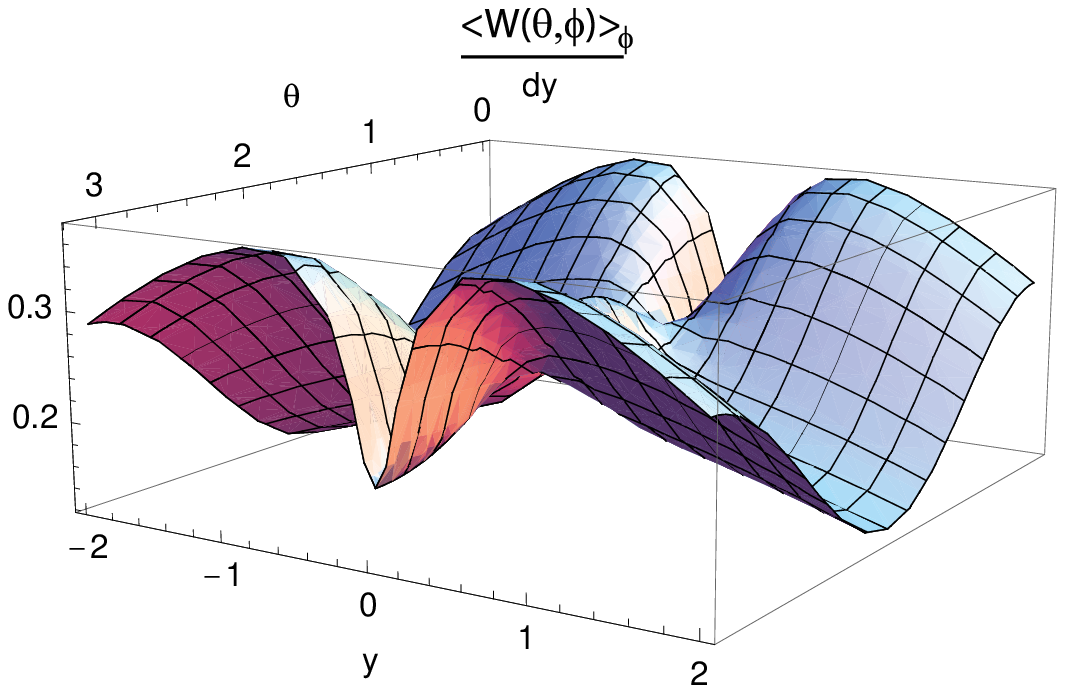}}
\end{minipage}
   \caption{\label{fig:Wav-y}
   \small \em Angular differential distribution of $J/\psi$ meson averaged over
 azimuthal angle $\phi$ as function of polar angle $\theta$ (in radians) and
 $\chi_c$ meson rapidity $y$ for $\chi_c(1^+)$ (left panel) and
 $\chi_c(2^+)$ mesons (right panel). KMR UGDF \cite{MR} was used.}
\end{figure}
%===================================================================

Now the main question is how to measure experimentally the
kinematical effects in the central exclusive $\chi_c(1^+,2^+)$
production reflected in the maxima/minima of rapidity distributions
$d\sigma^J_{\lambda\lambda}/dy$ for different meson helicities
$\lambda$. Kinematical effects under consideration are natural in
the helicity frame, in which we actually work\footnote{We are
thankful to S. Baranov for useful discussion of this point.}. Such a
frame can be used in a real experiment. In principle, analogous
calculations in other frames of reference can also be useful. As was
already mentioned above, the total (summed over all meson helicities
$\lambda$) cross section is regular around $y\sim0$ and does not
contain information about rapidity dependence of the partial helicity
contributions.

Instead, for the purpose of experimental identification of the
kinematical effects and separation of the meson helicity
contributions it is instructive to look at the correlation function
$d\langle W(\theta,\phi)\rangle_{\phi}/dy$ in both variables --
polar angle of $J/\psi$ meson and rapidity $y$ of $\chi_c$ meson
(or, equivalently, $J/\psi+\gamma$ system). We present such a
function for each $J=1,\,2$ meson in Fig.~\ref{fig:Wav-y}.
Maxima/minima in differential distributions
$d\sigma^J_{\lambda\lambda}/dy$ are then reflected in $y$-dependence
of the correlation function $d\langle
W(\theta,\phi)\rangle_{\phi}/dy$ at fixed $\theta$.

In principle, one can write formulae analogous to
Eqs.~(\ref{jpsi-1}) and (\ref{jpsi-2}) in another frame, where the
``peaks'' in rapidity near $y\sim0$ can be eliminated. But then one
looses some information about non-maximal helicity projections of
the production density matrix. Since helicity has different meaning
in different frames, our effect is purely of kinematical nature and thus
should also be different in different frames.

%------------------------------------------------------
\section{Conclusions and discussion}
%------------------------------------------------------

Our results can be summarized as follows:

We have calculated differential cross sections for central exclusive
$\chi_c(1^+,2^+)$ meson production for different spin polarisations
in the helicity frame. The integrated cross section for the maximal
helicity state is approximately an order of magnitude greater than
that for the non-maximal ones. We have shown that this effect has a
kinematical nature: heavy (non-relativistic, $y\to0$) meson is
dominantly produced in the $\lambda=0$ state, massless (relativistic,
$y\gtrsim 1$) meson is produced mostly in the $\lambda=\pm1$ states. In
the total cross section integrated over the meson rapidity $y$ the
second effect turns out to be dominated, i.e. relativistic with
$y\gtrsim 1$ $\chi_c$ mesons with maximal helicity ($\lambda=\pm1$
for $\chi_c(1^+)$ and $\lambda=\pm2$ for $\chi_c(2^+)$) are
preferably produced in the exclusive process $pp\to p\chi_c p$.

We used two different models for the $q_t$-dependent unintegrated
gluon distributions from the literature -- conventional KMR
distribution which includes Sudakov form factor \cite{MR} and the
nonlinear Kutak-Sta\'sto model \cite{KS05} based on unified
BFKL-DGLAP evolution. The contributions of the minimal helicity
state $\lambda=0$ given by the diagonal elements of the production
density matrix $\rho^{J=1,2}_{00}$ are close both $\chi_c(1^+,2^+)$
mesons and amount to $5-6$ \%. They are only weekly dependent on
UGDF (but rather strongly dependent on cuts in rapidity $y$). Such
quantities $\rho^{J=1,2}_{00}$ are, therefore, good
model-independent observables, which can be used to constrain the
underlying QCD diffractive production mechanism of heavy quarkonia.

We have calculated, in addition, angular distributions of $J/\psi$
meson $W(\theta,\phi)$ from radiative decays $\chi_c(1^+,2^+) \to
J/\psi + \gamma$ in the helicity frame, both analytically and
numerically. The function $W^J(\theta,\phi)$ contains an important
information about all independent elements of the production density
matrix $\rho^J_{\lambda\lambda'}$, both diagonal and non-diagonal.
Azimuthal angle $\phi$ dependence of outgoing $J/\psi$ mesons is
given by the real and imaginary parts of the non-diagonal elements
$\rho^J_{\lambda\lambda'},\,\lambda\not=\lambda'$, which are very
sensitive to the UGDF model used. A specific shape of the rapidity
dependence of the $J/\psi$ meson correlation function $d\langle
W^J(\theta,\phi)\rangle_{\phi}/dy$ shown in Fig.~\ref{fig:Wav-y} is
a result of the ``peaked'' shape of the $\chi_c$ production density
matrix around central rapidity region $y\sim0$. Measurements of such
a differential distribution $d\langle
W^J(\theta,\phi)\rangle_{\phi}/dy$ averaged over azimuthal angle
$\phi$ would allow to separate different helicity contributions and
to identify the kinematical effects under discussion. Distributions
based on general correlation function $W^J(\theta,\phi)$ can, in
principle, be measured at the Tevatron and LHC. They could also
provide an independent check of the discussed diffractive QCD
mechanism.

\begin{acknowledgments}
This work was partly supported by the Carl Trygger Foundation, the
RFBR (grants No. 09-02-01149 and 09-02-00732) and the Polish grant
of MNiSW No. N202 2492235. We are grateful to Sergey Baranov and
Valery Khoze for valuable and stimulating discussions. We are
especially indebted to Lucian Harland-Lang for the exchange of
Fortran codes, making possible a direct comparison of our calculations and to
find out the reasons of differences between them.
\end{acknowledgments}

%--------------------------------------------------------------------

%-----------------------------------------------------------------------


\begin{thebibliography}{100}

\bibitem{Albrow:2008pn}
  M.~G.~Albrow {\it et al.}  [FP420 R and D Collaboration],
  %``The FP420 R\&D Project: Higgs and New Physics with forward protons at the
  %LHC,''
  JINST {\bf 4}, T10001 (2009)
  [arXiv:0806.0302 [hep-ex]].
  %%CITATION = JINST,4,T10001;%%

\bibitem{Aaltonen:2009kg}
  T.~Aaltonen {\it et al.}  [CDF Collaboration],
  %``Observation of exclusive charmonium production and gamma+gamma to mu+mu- in
  %p+pbar collisions at sqrt{s} = 1.96 TeV,''
  Phys.\ Rev.\ Lett.\  {\bf 102}, 242001 (2009)
  [arXiv:0902.1271].

\bibitem{Albrow:2010yb}
  M.~G.~Albrow, T.~D.~Coughlin and J.~R.~Forshaw,
  %``Central Exclusive Particle Production at High Energy Hadron Colliders,''
  arXiv:1006.1289 [hep-ph].
  %%CITATION = ARXIV:1006.1289;%%

\bibitem{HarlandLang:2009qe}
  L.~A.~Harland-Lang, V.~A.~Khoze, M.~G.~Ryskin and W.~J.~Stirling,
  %``Central exclusive chi_c meson production at the Tevatron revisited,''
  Eur.\ Phys.\ J.\  C {\bf 65}, 433 (2010)
  [arXiv:0909.4748 [hep-ph]].
  %%CITATION = EPHJA,C65,433;%%

\bibitem{HarlandLang:2010ep}
  L.~A.~Harland-Lang, V.~A.~Khoze, M.~G.~Ryskin and W.~J.~Stirling,
  %``Standard candle central exclusive processes at the Tevatron and LHC,''
  arXiv:1005.0695 [hep-ph].
  %%CITATION = ARXIV:1005.0695;%%

\bibitem{KMR}
V.A. Khoze, A.D. Martin and M.G. Ryskin, Phys. Lett. B {\bf 401},
330 (1997);\\
V.A. Khoze, A.D. Martin and M.G. Ryskin, Eur. Phys.
J. C {\bf 23}, 311 (2002);\\
A.B. Kaidalov, V.A. Khoze, A.D. Martin
and M.G. Ryskin, Eur.\ Phys.\ J.\ C {\bf 31}, 387 (2003)
[arXiv:hep-ph/0307064];\\
A.B. Kaidalov, V.A. Khoze, A.D. Martin and M.G. Ryskin,
Eur. Phys. J. C {\bf 33}, 261 (2004);\\
V.A. Khoze, A.D. Martin, M.G. Ryskin and W.J. Stirling,
Eur. Phys. J. C {\bf 35}, 211 (2004).

\bibitem{chic1}
  R.~S.~Pasechnik, A.~Szczurek and O.~V.~Teryaev,
  %``Elastic double diffractive production of axial-vector \chi_c(1^{++}) mesons
  %and the Landau-Yang theorem,''
  Phys.\ Lett.\  B {\bf 680}, 62 (2009)
  [arXiv:0901.4187 [hep-ph]].
  %%CITATION = PHLTA,B680,62;%%

\bibitem{chic2}
  R.~S.~Pasechnik, A.~Szczurek and O.~V.~Teryaev,
  %``Nonperturbative and spin effects in the central exclusive production of
  %tensor \chi_c(2^+) meson,''
  Phys.\ Rev.\  D {\bf 81}, 034024 (2010)
  [arXiv:0912.4251 [hep-ph]].
  %%CITATION = PHRVA,D81,034024;%%

\bibitem{chic0}
  R.~S.~Pasechnik, A.~Szczurek and O.~V.~Teryaev,
  %``Central exclusive production of scalar \chi_c meson at the Tevatron, RHIC
  %and LHC energies,''
  Phys.\ Rev.\  D {\bf 78}, 014007 (2008)
  [arXiv:0709.0857 [hep-ph]].
  %%CITATION = PHRVA,D78,014007;%%

\bibitem{Lam-Tung}
  C.~S.~Lam and W.~K.~Tung,
  %``A Systematic Approach To Inclusive Lepton Pair Production In Hadronic
  %Collisions,''
  Phys.\ Rev.\  D {\bf 18}, 2447 (1978).
  %%CITATION = PHRVA,D18,2447;%%

\bibitem{Khoze:2000jm}
  V.~A.~Khoze, A.~D.~Martin and M.~G.~Ryskin,
  %``Double-diffractive processes in high-resolution missing-mass  experiments
  %at the Tevatron,''
  Eur.\ Phys.\ J.\  C {\bf 19}, 477 (2001)
  [Erratum-ibid.\  C {\bf 20}, 599 (2001)]
  [arXiv:hep-ph/0011393].
  %%CITATION = EPHJA,C19,477;%%

\bibitem{Saleev06}
  B.~A.~Kniehl, D.~V.~Vasin and V.~A.~Saleev,
  %``Charmonium production at high energy in the k(T)-factorization  approach,''
  Phys.\ Rev.\  D {\bf 73}, 074022 (2006)
  [arXiv:hep-ph/0602179].
  %%CITATION = PHRVA,D73,074022;%%

\bibitem{Kniehl03}
  B.~A.~Kniehl, G.~Kramer and C.~P.~Palisoc,
  %``chi/c1 and chi/c2 decay angular distributions at the Fermilab Tevatron,''
  Phys.\ Rev.\  D {\bf 68}, 114002 (2003)
  [arXiv:hep-ph/0307386].

\bibitem{Pilk} H. Pilkuhn, {\it The Interactions of Hadrons},
(North-Holland, Amsterdam, 1967), p.~222.

\bibitem{Sakur} J. J. Sakurai, {\it Modern Quantum Mechanics},
(Addison-Wesley, Reading, 1994), p.~223.

\bibitem{e835} E835 Collaboration, M. Ambrogiani {\it et al.},
Phys.\ Rev.\ D {\bf65}, 052002 (2002).

\bibitem{Khoze:2002nf}
  V.~A.~Khoze, A.~D.~Martin and M.~G.~Ryskin,
  %``Physics with tagged forward protons at the LHC,''
  Eur.\ Phys.\ J.\  C {\bf 24}, 581 (2002)
  [arXiv:hep-ph/0203122].
  %%CITATION = EPHJA,C24,581;%%

\bibitem{KS05}
K. Kutak and A.M. Sta\'sto, Eur. Phys. J. {\bf C41}, 341 (2005).

\bibitem{MR}
  M.~A.~Kimber, A.~D.~Martin and M.~G.~Ryskin,
  %``Unintegrated parton distributions,''
  Phys.\ Rev.\  D {\bf 63}, 114027 (2001)
  [arXiv:hep-ph/0101348];\\
  %%CITATION = PHRVA,D63,114027;%%
  A.~D.~Martin and M.~G.~Ryskin,
  %``Unintegrated generalised parton distributions,''
  Phys.\ Rev.\  D {\bf 64}, 094017 (2001)
  [arXiv:hep-ph/0107149].
  %%CITATION = PHRVA,D64,094017;%%

\bibitem{GRV}
  M. Gl\"uck, E. Reya and A. Vogt, Z. Phys. C {\bf 67}, 433 (1995);\\
  M. Gl\"uck, E. Reya and A. Vogt, Eur. Phys. J. C {\bf 5}, 461 (1998).


\bibitem{qcdcorr}
  R.~Barbieri, M.~Caffo, R.~Gatto and E.~Remiddi,
  %``QCD Corrections To P Wave Quarkonium Decays,''
  Nucl.\ Phys.\  B {\bf 192}, 61 (1981);\\
  %%CITATION = NUPHA,B192,61;%%
  W.~Kwong, P.~B.~Mackenzie, R.~Rosenfeld and J.~L.~Rosner,
  %``Quarkonium Annihilation Rates,''
  Phys.\ Rev.\  D {\bf 37}, 3210 (1988);\\
  %%CITATION = PHRVA,D37,3210;%%
  M.~L.~Mangano and A.~Petrelli,
  %``An update on chi($c$) decays: Perturbative QCD versus data,''
  Phys.\ Lett.\  B {\bf 352}, 445 (1995)
  [arXiv:hep-ph/9503465].
  %%CITATION = PHLTA,B352,445;%%


\end{thebibliography}
\end{document}